\documentclass[12pt]{article}

\usepackage{amsmath,amssymb} 

\usepackage{epsf}
\usepackage{bbm}
\usepackage{pstricks}
\usepackage[pdftex]{graphicx}
\usepackage{epstopdf}
\usepackage{cite}

\newcommand{\beq}{\begin{eqnarray}}
\newcommand{\eeq}{\end{eqnarray}}

\newcommand{\centeron}[2]{{\setbox0=\hbox{#1}\setbox1=\hbox{#2}\ifdim

\wd1>\wd0\kern.5\wd1\kern-.5\wd0\fi \copy0

\kern-.5\wd0\kern-.5\wd1\copy1\ifdim\wd0>\wd1
                                      \kern.5\wd0\kern-.5\wd1\fi}}
\newcommand{\ltap}{\>\centeron{\raise.35ex\hbox{$<$}}
                              {\lower.65ex\hbox{$\sim$}}\>}
\newcommand{\gtap}{\>\centeron{\raise.35ex\hbox{$>$}}
                              {\lower.65ex\hbox{$\sim$}}\>}

\newcommand\ZZ{\hbox{\zfont Z\kern-.4emZ}}
\font\zfont = cmss10 

\newcommand{\vev}[1]{{\langle #1 \rangle}}

\def\lt{\left}
\def\rt{\right}
\def\ie{{\it i.e.\ }}
\def\eg{{\it e.g.\ }}
\def\ord{{\cal O}}
\def\beqa{\begin{eqnarray}}
\def\eeqa{\end{eqnarray}}
\def\half{\frac{1}{2}}
\def\ben{\begin{enumerate}}
\def\een{\end{enumerate}}
\def\bei{\begin{itemize}}
\def\eei{\end{itemize}}

\def\epu{{\epsilon_u}}
\def\NKK{{N_{\rm KK}}}
\def\mKK{{m_{\rm KK}}}

\newcommand{\be}{\begin{eqnarray}}
\newcommand{\ee}{\end{eqnarray}}
\newcommand{\bpm}{\begin{pmatrix}}
\newcommand{\epm}{\end{pmatrix}}

\begin{document}
\begin{titlepage}



\vspace*{-2.0cm}
\begin{center}
{\huge \bf Flavor Alignment via Shining in RS}
\end{center}
\vskip0.2cm

\begin{center}
{\bf Csaba Cs\'aki$^a$, Gilad Perez$^b$,
Ze'ev Surujon$^c$ and Andreas Weiler$^d$}

\end{center}
\vskip 4pt

\begin{center}
$^{a}$ {\it Institute for High Energy Phenomenology\\
Newman Laboratory of Elementary Particle Physics\\
Cornell University, Ithaca, NY 14853, USA } \\
\vspace*{0.3cm}
$^b$ {\it Department of Particle Physics
Weizmann Institute of Science\\
Rehovot 76100, Israel} \\
\vspace*{0.3cm} $^c$ {\it Department of Physics\\
University of California at San Diego\\
La Jolla, CA 92093, USA}\\
\vspace*{0.3cm} $^d$ {\it CERN Theory Division\\
CH-1211 Geneva 23, Switzerland} \\

\vspace*{0.1cm}

{\tt  csaki@cornell.edu,
gilad.perez@weizmann.ac.il, zevik@ucsd.edu,
andreas.weiler@cern.ch}
\end{center}

\vglue 0.3truecm

\begin{abstract}
\vskip 3pt \noindent We present a class of warped extra dimensional
models whose flavor violating interactions are much suppressed
compared to the usual anarchic case due to flavor alignment.
Such suppression can be achieved in
models where part of the global flavor symmetry is gauged in the bulk
and broken in a controlled manner.
We show that the bulk masses can be aligned with the down type Yukawa couplings
by an appropriate choice of bulk flavon field representations and TeV brane dynamics.
This alignment could reduce the flavor
violating effects to levels which allow for a Kaluza-Klein scale as
low as 2-3~TeV, making the model
observable at the LHC.
However, the up-type Yukawa couplings on the IR brane, which are bounded from below by recent bounds on CP violation in the $D$ system,
induce flavor misalignment radiatively.
Off-diagonal down-type Yukawa couplings and kinetic mixings for the down quarks are both consequences of this effect.
These radiative Yukawa corrections can be reduced by raising the flavon VEV on the IR brane (at the price of some moderate tuning), or by extending the Higgs sector.
The flavor changing effects from the radiatively induced Yukawa mixing terms are at
around the current upper experimental bounds. We also show the
generic bounds on UV-brane induced flavor violating effects, and
comment on possible additional flavor violations from bulk flavor
gauge bosons and the bulk Yukawa scalars.
\end{abstract}

\end{titlepage}



\section{Introduction}
\label{sec:Intro} \setcounter{equation}{0} \setcounter{footnote}{0}

Efforts to solve the hierarchy problem of the Standard Model (SM)
usually end up introducing new physics at the TeV scale. However,
most of these models would also allow new flavor violation effects
which have not been observed.
This usually leads one to require that the new physics is very nearly
flavor universal, which would exclude the possibility of finding
a testable solution to the fermion mass hierarchy problem within the same
framework.

Warped extra dimensions~\cite{RS1} offer new approaches to flavor
physics~\cite{GN}. The puzzle of the hierarchical flavor
structure can be solved via the split fermion mechanism~\cite{ArkaniHamed:1999dc}, using
flavor dependent wave functions for the Standard Model fermions~\cite{RSoriginal}.
This "anarchic" approach to flavor naturally yields the hierarchies
of masses and mixing angles.
Within this framework, the same setup also induces extra protection against
excess of flavor changing neutral current (FCNC) processes~\cite{Huber:2003tu,aps} (see also~\cite{Burdman:2002gr}), as
there is a built-in RS-GIM mechanism~\cite{aps} where flavor violation is dominantly
due to non-universality of fermion profiles near the IR brane.
Thus, flavor and CP violation (CPV) related to the first two generations is highly suppressed~\cite{aps,NMFV} as in the SM case.
A  residual little CP problem is, however, still found in the form of too large contributions
to the neutron electric dipole moment (EDM)~\cite{aps}, and sizable chiraly enhanced contributions to
$\epsilon_K$~\cite{cfw1,Blanke:2008zb,Davidson:2007si,UTFit,rsflavorrecent}.
FCNCs are dominated
by the exchange of Kaluza-Klein (KK) gluons, which induce the most
dangerous (LR-type) $\Delta F=2$ four-Fermi operators contributing
to $K-\bar{K}$ mixing. The lower bound on the KK mass scale (taking
into account the effect of the RS-GIM mechanism) is, at leading order, around  $20$~TeV~\cite{cfw1} (compared with $\sim 10^4-10^5$~TeV~\cite{UTFit} for
flavor-structureless models), which would render the model
unobservable at the LHC.
Recently
in~\cite{Agashe:2008uz} it was argued, based on matching the full
RS set-up onto a two site model, that if the Higgs is in the bulk
and one loop matching of the gauge coupling is included, the KK
scale can be lowered down to around 5-6 TeV.
The weaker bound is due to a combination of several effects:
(i) one loop matching of the gauge coupling which lowers the leading order bound by a factor of two, down to roughly 10 TeV, above the LHC reach~\cite{KKgluon};
(ii) exploiting effects related to the presence of a bulk Higgs which amounts to effectively increasing the overall scale
of the 5D Yukawa coupling, allowing for more elementary light fermions.
In the analysis below, we shall apply the one loop matching result, item (i). Increasing the size of the 5D Yukawa however,
item (ii), implies that RS chirality violating processes such as the ones which contributes to EDMs, $b\to s\gamma$~\cite{aps} are enhanced.
In particular, it was recently shown that the bound from $\epsilon'/\epsilon_K$~\cite{GIP} constrains the KK scale to be above the ${\cal O}(6)\,$TeV
scale, still above the LHC reach (see also~\cite{HFCNC} for Higgs mediated flavor violation processes which are also enhanced in this limit).
Thus, in our study we assume an IR brane localized Higgs, and our results can be viewed as conservative ones
in the sense that various constraints might be further ameliorated with a bulk Higgs.

The little CP problem might be accidentally solved if a
combination of various \emph{unrelated} CP violating parameters just happen to be small, of ${\cal O}(0.1)$.
However, in view of the fact that the CKM phase is of ${\cal O}(1)$ we are motivated to look for a parametric solution
for the problem.
One may seek guidance by using the AdS/CFT correspondence~\cite{AdSCFT},
where the RS setup can be understood from a 4D point of view~\cite{AdSCFTpheno}, identifying 4D global
currents with bulk gauge fields.
In fact, in the context of electroweak precision tests, progress was made~\cite{Agashe:2003zs}
by gauging the approximate custodial symmetry of the SM in the bulk, as motivated by the duality.
Thus, we expect that a similar progress can be made by gauging (part or all of) the SM approximate flavor symmetries~\cite{Rattazzi,xdgim,FPR,cfw2,LMFV,Csaki:2008qq,muchun,santiago,Csaki:2009bb}.

There are several directions of how to realize models with gauged 5D flavor symmetries.
One possible distinction is whether flavor issues (the flavor puzzle and the flavor problem) are addressed solely by Planck physics on the UV brane or whether bulk physics participates in the flavor dynamics as well.
Obviously, our ability to directly and indirectly probe flavor dynamics
in the near future depends on which of these cases is realized in nature.
In~\cite{Rattazzi,xdgim,Csaki:2009bb} it was proposed to break the flavor symmetries only on the UV brane.
 Such a setup leads to a class of models where the flavor puzzle is solved by Planck scale physics
and described by 4D general minimal flavor violation (GMFV)~\cite{MFV,GMFV} (where the 4D Yukawa matrices control the flavor
violation but generically one needs to resum the contributions related to the third generation fermions~\cite{GMFV}).
This class of models is in general similar in structure to the SM, hence hard to probe via flavor precision tests.

Another class of models which was recently proposed~\cite{FPR,cfw2,santiago} (see also~\cite{LMFV,Csaki:2008qq,muchun} for the lepton sector)
try to solve the little CP problem but without giving up on addressing the flavor puzzle (and taking advantage on the built-in RS-GIM).
These models may allow us to directly probe new flavor dynamics
at the the LHC, which is very exciting, as well as leading to deviate from the
SM predictions in indirect flavor precision tests.
The basic idea is to align the down type quark sector which includes the bulk masses and the
5D down Yukawas (and possibly also the brane kinetic terms) such that the constraint from $\epsilon_K$
is satisfied.
The model proposed in~\cite{cfw2} is based on combining two U(1) horizontal symmetries broken on the UV and IR branes respectively
and an additional flavor diagonal U(1) symmetry (motivated by the custodial symmetry to $Z\to b\bar b$~\cite{newcustodial})
which allows one to split-up the SM weak doublet partners into two separate 5D multiplets.
In~\cite{santiago} a global U(3) symmetry for the down quark weak singlet fields was proposed
in order to provide the required alignment.\footnote{Since it is expected that the bulk
physics, which include quantum gravity effects, would generically violate any global symmetries, we imagine that this symmetry is actually gauged.}
In this model one has to assume a mild hierarchy in the 5D down Yukawa parameters, and there is also a tension related to mixing with KK states due to the absence of RS-GIM protection for the RH down type quarks~(see~\cite{LMFV} for a related discussion).
In~\cite{FPR}, 5D MFV (5DMFV) was proposed as a framework which may allow to control
the form of flavor violation in warped models.
Generic 5DMFV models at low energies yield a flavor structure which is similar to generic RS models,
and hence do not solve the little CP problem~\cite{FPR}. However, 5DMFV allows to ameliorate the problem
via accidental approximate alignment: a consequence of
a single flavor diagonal parameter, $r_u$, which controls the tree level down quark flavor violation, being accidentally small.
As discussed below, however, the RS $\epsilon_K$ problem requires a rather high degree of alignment, $r_u={\cal O}(10^{-2})$, which calls for a non-accidental
solution.

In this work we examine how to combine aspects of 5DMFV and alignment in order to
construct a viable class of models in which the assumptions on the UV (Planckian) breaking
of flavor are minimal.
An important tool which we use here is the mechanism of
``shining''~\cite{shining}, \ie transmitting a symmetry breaking
effect from the UV brane through the bulk by scalar fields.
This idea was first applied
to warped space flavor models in~\cite{Rattazzi} to explain how
natural flavor conservation can appear in an RS type model.
In~\cite{Rattazzi}, the fermions were assumed to be confined to the
IR brane, so in that case one could protect the fermions from flavor
violation but there was no explanation for the flavor hierarchy.
The main new ingredient here is that we are also allowing the field
which transmits the flavor violation to
couple to the bulk fermions, resulting in small flavor violating
effects in the bulk via higher dimensional operators. These effects are then manifested
as non universal bulk
masses for the quarks, giving rise to flavor dependent wave functions for
the quark zero modes, like in~\cite{FPR}, thus explaining the flavor
hierarchy.

We will present two different symmetry breaking patterns:
In the first model, we impose a SU(3)$_Q\times$SU(3)$_d$ flavor symmetry,
while in the second one we impose only the diagonal subgroup SU(3)$_{Q+d}$.
In both cases we assign the bulk scalars with the quantum numbers of the
down-type Yukawa coupling $Y_d$, namely $({\bf 3},{\bf \bar 3})$ in the first case, and ${\bf 8}$
in the second.
We do not impose any symmetry in the right-handed up sector,
thus allowing for anarchic bulk masses, but we require that the up-type Yukawa
coupling is an IR brane field, so that it does not feed into VEVs of other bulk fields.
This approach actually leads to the fact that the up type sector tends to be anarchic,
hence (as in other alignment models - see {\it e.g}~\cite{Nir:2002ah,cfw2,FPR} and references therein), the presence of up type flavor and CPV is expected in our scenario.
Recently, however, it was pointed out~\cite{DCPV} that a strong bound from CPV in $D^0-\bar D^0$ mixing already exists
and is translated to a strict lower bound on the scale of the 5D up type Yukawa matrix in anarchic RS models.
This implies that the tree-level alignment is going to be spoiled due to higher order effects induced by the sizable
up type Yukawa matrix. We will examine these misaligning effects in great detail, and show possible ways of reducing their effects below the current experimental bounds. We also comment on possible flavor effects of the bulk scalars and flavor gauge bosons.

The paper is organized as follows: in section~\ref{sec:setup} we give a basic realization of 5DMFV,
and show that in order to have alignment, one needs a more restricted symmetry breaking pattern
than the generic 5DMFV of~\cite{FPR}.
In section~\ref{sec:models}, we present the two improved models in which the down type Yukawa is
aligned with the bulk quark masses.
In section~\ref{sec:loops} we discuss misalignment effects beyond the leading order.
These include effects at the IR brane due to the presence of the
up-type Yukawa background field, effects of both IR and UV brane kinetic terms. We comment on possible FCNCs from flavons and flavor gauge fields.
We conclude in section~\ref{sec:conclusion}.

\section{The Setup }
\label{sec:setup} \setcounter{equation}{0}
\setcounter{footnote}{0}

First we construct a model where bulk scalars transmit (shine)
flavor violation through the bulk, giving a concrete realization to
5DMFV. We then show that the suppression of flavor violating effects
in generic 5DMFV models is quite a bit less than originally
expected, motivating us to look for models with complete alignment.

\subsection{Generic 5DMFV from Shining without Alignment}
We start with some definitions and notations, which will be used throughout this paper.
Recall that using conformally flat coordinates, the RS metric is given by a slice of AdS$_5$
\begin{equation}  \label{eq:metric}
ds^2=\left(\frac{R}{z}\right)^2 (dx_\mu dx_\nu \eta^{\mu\nu} -dz^2),
\end{equation}
bounded by a UV brane at $z=R$ ($R$ is also the AdS curvature scale) and an
IR brane at $z=R'$.
The magnitudes of the scales are given by $R^{-1}\sim M_{Pl}$ and $R'^{-1} = 1-2$ TeV.

The electroweak gauge group is extended to
SU(2)$_L\times$SU(2)$_R\times$U(1)$_{B-L}$ gauge symmetry in the
bulk to incorporate custodial symmetry~\cite{Agashe:2003zs}. This symmetry is reduced on
the UV brane to the SM group SU(2)$_L\times$U(1)$_Y$. We assign the
following representations for the quarks: \beq
   Q_i(2,1,1/3),\quad u_i(1,2,1/3),\quad d_i(1,2,1/3)\qquad (i=1,2,3),
\eeq so that there are two separate SU(2)$_R$ doublets per
generation.

We impose the flavor symmetry SU(3)$_Q\times$SU(3)$_u\times$SU(3)$_d$ in the bulk.
The improved models discussed in section \ref{sec:models} have only a subgroup
of this as their bulk global symmetry.
Gauging this group will give rise to flavor gauge bosons which mediate FCNCs once
the symmetry is broken.
We will also assume that the only sources for flavor violation are
on the branes. In the general case, these will break the flavor
symmetry in the UV completely (we will later show the bounds on UV
flavor violating effects). The UV flavor violation is then
transmitted (shined) through the bulk via bulk scalars which couple
to the UV flavor sources and acquire VEVs.

In order to transmit the flavor violation, we introduce bulk scalar fields $y_{u,d}$ 
which transform under the flavor group (the precise representation of the bulk fields is model dependent and  is discussed below).
These scalars will be the origin of the brane localized Yukawa
couplings, so that the 5D Yukawa terms then become
\beq
   \lambda_u y_{u\,ij}\bar Q_i \tilde H u_j+   \lambda_d y_{d\,ij}\bar Q_i H d_j,
\eeq
These scalars $y_{u,d}$ are assumed to have small bulk masses compared to the
AdS curvature,  and to first approximation their expectation values
are uniform (it is easy to extend the discussion to non-constant
VEVs, see the appendix).
 For example, consider a simple case,  where the bulk fields, $y_{u,d}$, are bi-fundamental
of the 5D flavor group, so that in our notation the usual dimensionless Yukawas are 
$Y_{u,d}=
\lambda_{u,d}\vev{y_{u,d}}R^{3/2}$.
The main difference between this approach and that of~\cite{Rattazzi}
is that we are also considering the effects of small higher
dimension operators coupling the bulk fermions to the flavor
symmetry breaking bulk scalars. This coupling will be the source of
the flavor hierarchy. Although the coefficients of these operators
are suppressed by the cutoff scale, their effect can still be very
significant, due to the exponential sensitivity of the effective 4D
masses on the bulk mass parameters.

\subsubsection*{Effective Bulk Fermion Mass}
To order $Y^3$, the effective bulk fermion
masses (in units of the AdS curvature) are of the form
\begin{eqnarray}    \label{eq:bulk-masses}
c_Q &=& \alpha_Q \cdot 1 + \beta_Q Y_d Y_d^\dagger + \gamma_Q
Y_u Y_u^\dagger \\
c_u &=& \alpha_u \cdot 1 + \gamma_u Y_u^\dagger Y_u\\
c_d &=& \alpha_d \cdot 1 + \beta_d Y_d^\dagger Y_d.
\end{eqnarray}
Here the $\alpha_i$ are flavor-invariant bulk mass parameters which
are $\ord(1)$. These can not be forbidden by any flavor symmetry.
Although this term was omitted in~\cite{FPR}, it does play an
important role both for producing the leading order operator
(compared to which the insertions of the bulk scalars are considered
as perturbations), and for getting the desired alignment in flavor
space.

The parameters $\beta_i, \gamma_i$ are coefficients of
higher dimension operators suppressed by the cutoff scale $\Lambda$.
Let us estimate how large $\beta , \gamma$ could be.
The corresponding effective operators are of the form \eg $B y_u^\dagger y_u \bar{Q} Q$,
where $B$ is a coefficient of dimension $-2$.
Naive Dimensional Analysis (NDA) then suggests that $B \lesssim (4\pi/\Lambda)^2$, leading
to $\beta \equiv B /R^2 \lesssim (4\pi/\Lambda R)^2$,
which may be a significant correction to the leading term.
In order to get a reliable prediction for the bulk masses $c_i$, one would also
expect that the correction to the universal term after the insertion
of the VEVs is small, so that the next correction is truly
negligible numerically.
  However, the improved models we will present in the next section exhibit
complete alignment in the bulk, so that while a higher order term
would change the numerical values of the expansion coefficients, it
would not spoil the alignment itself.

\subsubsection*{Flavor Changing Neutral Currents}

Given Eq.(\ref{eq:bulk-masses}), the 4D mass matrices are given by
\begin{eqnarray}
m^{(u)}_{ij} &=&
\frac{v}{\sqrt{2}} F_Q Y_u F_u, \nonumber \\
m^{(d)}_{ij} &=&
\frac{v}{\sqrt{2}} F_Q Y_d F_d,
\end{eqnarray}
where we have used the normalized IR values of the quark zero-mode wave functions, $f_{Q,u,d}$~\cite{aps},
as being the eigenvalues of the spurions $F_{Q,u,d}$ (in the basis where the bulk masses, $c_{Q,u,d}$, are diagonal)
\begin{equation}
f_{x} = \sqrt\frac{1 -2 c_x}{1-
(\frac{R}{R'})^{1-2c_x}},
\end{equation}
Let us show explicitly the
result of~\cite{FPR} that in generic GMFV models tree-level FCNC's are similar in rate to the anarchic case. For
concreteness we focus on the down-type quark sector (since these processes are the ones most
severely constrained, though the analysis is completely analogous
for the FCNC's involving up-type quarks).
We use the bulk U(3)$^3$ flavor symmetry to diagonalize the matrix
$Y_d$, which is achieved by a bulk basis transformation of the $Q_L$
and the $d_R$.
We can perform a further bulk unitary transformation on
$u_R$, after which the form of the bulk Yukawas is
\begin{equation}
Y_u = V_5 U^{(diag)}, \ \ Y_d = D^{(diag)}.
\end{equation}
Note that since $Y_u$ feeds into the bulk mass
parameter of $Q_L$, diagonalizing $Y_d$ is {\em not} sufficient
to diagonalize the down-type mass matrix.
In fact, at this point, $c_Q$ is not diagonal (while $c_{u,d}$ are).
The $c_Q$ can be diagonalized by $Q_L \to U_Q Q_L$.
Now one can calculate the zero mode wave
functions for these bulk fermions, and the effective down-type brane
Yukawa coupling takes the form
\begin{equation}
\bar{Q}_{L\ i} {f_Q}_i {U_Q^\dagger}_{ij} D^{(diag)}_{jk} {f_d}_k
{d_R}_k
\end{equation}
To diagonalize this matrix we need an additional rotation on the 4D
zero mode fields $Q_L\to V_Q Q_L, \ d_R \to V_d d_R$, such that
$v\times V_Q^\dagger {F_Q} {U_Q^\dagger} D^{(diag)} {F_d} V_d = {\rm diag}
(m_d,m_s,m_b)$.
The origin of FCNC's is then the following: with
diagonal bulk mass terms the fermion couplings to neutral gauge
bosons is diagonal, but not universal, due to the different wave
functions of the fermions. Then rotating the non-universal diagonal
coupling matrices to the actual 4D mass basis will induce flavor
changing couplings.
The leading effect comes from the first KK mode of the gluon,
\begin{equation}
g_{s*} \left[ \bar{Q}_L V_Q^\dagger F_Q \gamma_\mu \gamma (c_Q) F_Q
V_Q Q_L + \bar{d}_R V_d^\dagger F_d \gamma_\mu \gamma (c_d) F_d V_d
d_R \right] G^{(1)\mu},
\end{equation}
where $\gamma (c)\approx \frac{\sqrt{2} x_1}{J_1(x_1)}
\frac{0.7}{6-4 c}$, and $g_{s*}$ is the bulk gauge coupling in units
of the AdS curvature.
Therefore, indeed the generic 5DMFV framework contains four-Fermi operators of the form
$\bar{d}_{L,R} \gamma_\mu d_{L,R} \bar{d}_{L,R} \gamma^\mu d_{L,R}$.
The most dangerous of
these is the LLRR, or $(V-A)(V+A)$ operator, which is absent from the SM and has to be suppressed which requires some form of alignment
to make the model viable~\cite{FPR}.

\subsection{Approximate Alignment}
Let us now consider the limiting cases when the left handed bulk
mass only couples to either up or down type flavor:
\begin{equation}\label{5DGIM}
c_Q = \alpha_Q \cdot 1 + r_u \beta_Q Y_uY_u^\dagger + r_d \gamma_Q
Y_d Y_d^\dagger
\end{equation}
in the limit of $r_u\to 0$ or $r_d\to 0$.
Consider the case $r_u=0$.
In this case we choose a bulk basis where
$Y_d$ is diagonal, so that the bulk masses for the $Q_L$ fields and
for $d_R$ are automatically diagonal.
Then the 4D mass matrix for the down type quarks is automatically diagonal in the same basis
where the couplings to the bulk gauge fields are diagonal, so that
there are no treel-level FCNCs involving down-type quarks.
The mass matrix for the up-type quarks is not diagonal, and the misalignment between the two directions in flavor space implies the presence of up-type quark FCNCs.
Since in this case the bulk mass matrices are aligned with the down-type Yukawa
coupling, flavor violations occur only in the up sector.
The case analyzed in~\cite{FPR} corresponds to $r_u \ll r_d$, where down-type quark FCNCs
are naively suppressed by $r_u^2$.
However, this is not always the case: the
suppression also depends on the values of the 5D CKM mixing
angles, and usually the actual suppression is quite a bit less than
the naive $r_u^2$.
The origin of this effect can be understood as follows:  due to the 5D GIM mechanism the universal part in $c_Q$ has no flavor content and it cannot
induce large CP breaking required to produce the CKM order one phase (this is just a manifestation of the 5D Kobayashi-Maskawa mechanism). On the same lines
the suppression in FCNCs is proportional to the alignment of the {\it traceless} parts of $c_Q$ and $Y_d^\dagger Y_d$.
The actual suppression therefore depends on the amount of deviation from degeneracy between the
various eigenvalues of the $r_u$  and $r_d$ terms in Eq. (\ref{5DGIM}).
An explicit example for this which can be treated analytically is the two flavor case,
which we present below.
Then we proceed to a numerical analysis of the suppression in the three flavor case.

\subsubsection*{FCNC Suppression in Two Generations}

In the two flavor case, the 5D Yukawa couplings can be parameterized as
\begin{eqnarray}
     Y_u &=&  V_5 Y_u^{\rm diag} =  \,\begin{pmatrix}   \cos \theta & \sin \theta\\ -\sin \theta & \cos \theta\end{pmatrix}\begin{pmatrix}  y_{u,1} & 0 \\ 0 & y_{u,2}\end{pmatrix}, \\
    Y_d &=&  \,\begin{pmatrix}  y_{d,1} & 0 \\ 0 & y_{d,2}\end{pmatrix},
    \end{eqnarray}
where $V_5 = V_5(\theta)$ is the two flavor 5D CKM
    matrix.  Just like in the SM with two flavors, CP is conserved and the Yukawas are real. The bulk mass $c_Q$ is given by
    \begin{eqnarray}
    c_Q &=&  \alpha_Q \, \mathbf{1} + \beta^L_d \, Y_d Y_d^\dagger  + \beta^L_u \, Y_u Y_u^\dagger
    ,\\
     &=& \alpha_Q \,  \mathbf{1}  + \beta^L_d\,
   {\rm diag} \left(  y_{d,1}^2, y_{d,2}^2 \right)   +  \beta^L_u \, V_5 \,    {\rm diag} \left( {y_{u,1}^2, y_{u,2}^2}\right)     V_5^T.
\end{eqnarray}
In this basis, $c_d$ and $c_u$ are already diagonal.
Off-diagonal elements of $c_Q$ signal tree-level FCNCs in the down-type quark
sector.
Let us introduce the matrix $U_Q=U_Q(\phi)$ which diagonalizes $c_Q$.
%
%
The angle $\phi$ parametrizes the misalignment between $c_Q$ and the $Y_d$, and measures
the size of FCNCs in the down sector.
We can find $U_Q$ by solving
\begin{eqnarray}
    U_Q^T \left\{{\rm diag} \left( {c_Q^1,c_Q^2}\right) \right \}U_Q  &\stackrel{!}{=}&    c_Q - \frac12 {\rm tr} [c_Q] \nonumber\\
    &=&
   \frac12 \beta^L_d \left(
      y_{d,1}^2 - y_{d,2}^2 \right)
     \sigma_3  + \frac12 \beta^L_u \left(
         y_{u,1}^2-  y_{u,1}^2 \right) V_5   \sigma_3    V_5^T,
\end{eqnarray}
where we have included only the traceless contributions to $c_Q$
since  terms proportional to the identity do not affect $U_Q(\phi)$.
In the limit of small misalignment $(\sin \phi \ll 1)$ we find that
\begin{eqnarray}
    \sin\phi = \kappa \cos\theta \, \sin\theta + \mathcal{O}(\kappa^2),
\end{eqnarray}
where
\begin{eqnarray}\label{twoDkappa}
     \kappa \equiv \frac{\beta^L_u}{\beta^L_d} \frac{
         \left(y_u^1\right)^2-  \left(y_u^2\right)^2 }{
         \left(y_d^1\right)^2 - \left(y_d^2\right)^2}
\end{eqnarray}
is the parameter relevant for the suppression of tree-level FCNCs,
assuming no fine-tuning of the 5D CKM angle $\theta$.

As expected we find that tree-level FCNCs are suppressed if the
the contribution of the traceless part of $Y_u Y_u^\dagger$ to $c_Q$ is small
compared with that of the traceless part of $Y_d Y_d^\dagger$.

\subsubsection*{FCNC Suppression in Three Generations}
\begin{figure}
\begin{center}
\includegraphics[width=4.5cm]{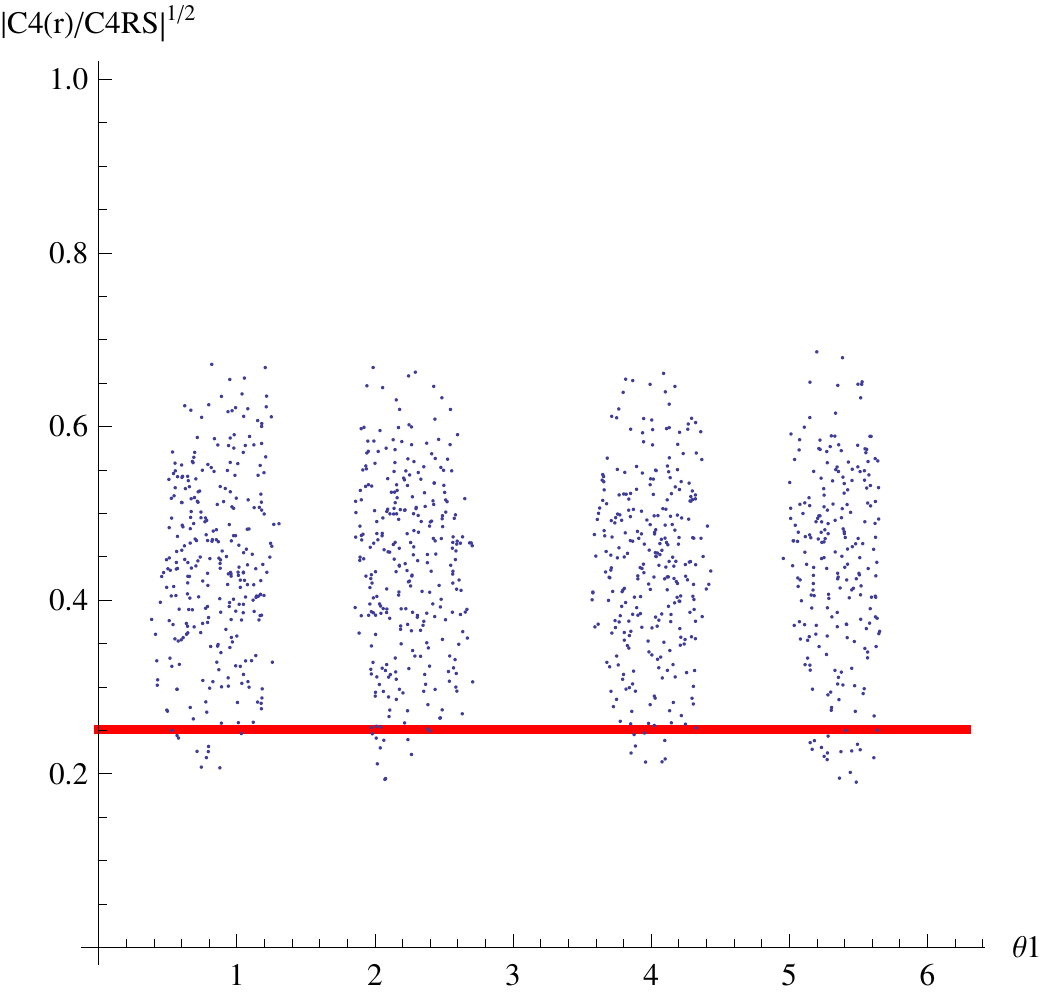}\quad
\includegraphics[width=4.5cm]{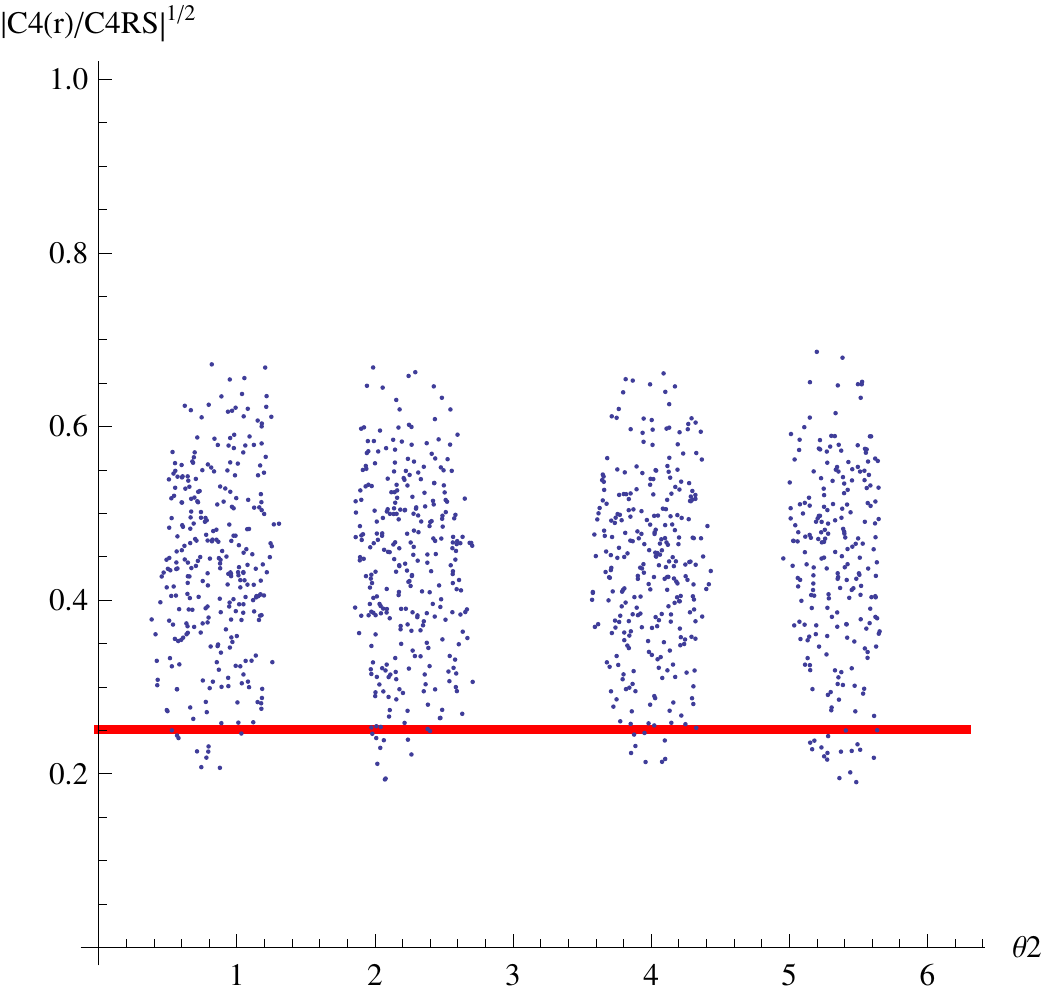}\quad
\includegraphics[width=4.5cm]{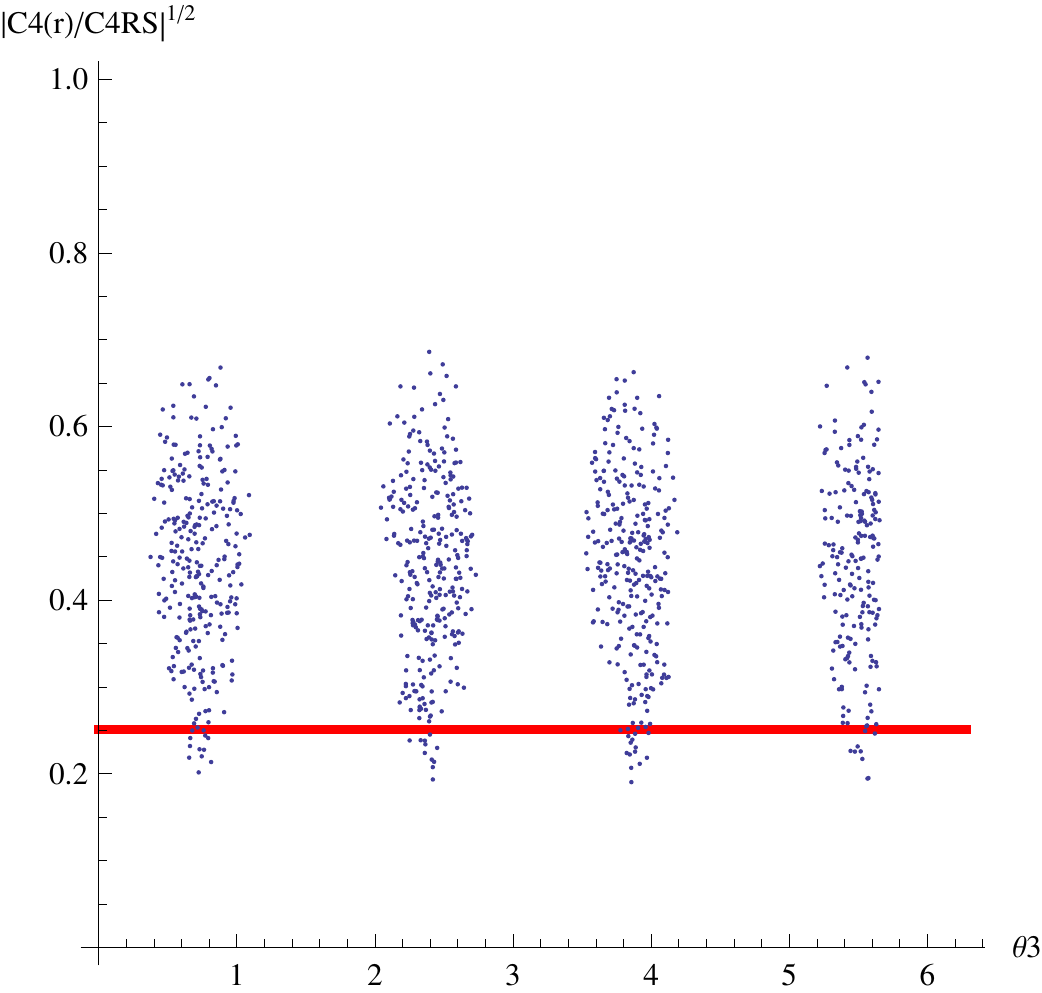}
\end{center}
\caption{We show the ratio of Wilson coefficients $\left|C_4^{\rm
5D-MFV}/C_4^{\rm RS}\right|^\frac{1}{2}$ vs. the naive expectation
$r  = 0.25$ (red line) as a function of the 5D CKM angles
$\theta_1,\theta_2$, and $\theta_3$. We included only points that
result in approximately the correct CKM matrix and Jarlskog determinant.\label{5dmfvscan} }
\end{figure}

We can use the analytical insights of the previous section to
understand the more complicated three-flavor case. Unfortunately, no
simple formula is available which relates the misalignment to the
coefficients and Yukawa eigenvalues.
However, we have learned that
$r=\beta^L_u/\beta^L_d$ alone is not a good predictor for small
FCNCs.
Certainly in the limit $r\rightarrow 0$, FCNCs vanish.
Due to the other factor involving the ratio of traceless contributions, the suppression can be weaker than $\sim r^2$. We will now argue  that in fact we generically expect
less suppression because the ratio of traceless contributions to
$Y_u Y_u^\dagger$ and $Y_d Y_d^\dagger$ is usually larger than one.
The size of the respective traceless contributions  is determined by
the size of the non-degeneracy of of $c_u$ and $c_d$. Due to the
hierarchies in the CKM and the mass spectrum~\cite{Xing:2007fb},
where
\beq
   \frac{m_d}{m_s \lambda_C}\sim 0.2, \quad \frac{m_d}{m_b \lambda_C^3}\sim 0.09,\quad
   \frac{m_u}{m_c \lambda_C}\sim 0.009, \quad \frac{m_u}{m_t \lambda_C^3}\sim 0.0006,
\eeq
and also
 since the down quark singlet masses are in the (exponentially suppressed) region where $c_d>0.5$~(see {\it e.g.}~\cite{aps} for a detailed discussion)
 the degeneracy in $c_d$ is much stronger than in $c_u$ (see section \ref{sec:models}).
We therefore
generically expect the traceless contribution to $Y_u Y_u^\dagger$
to be larger than the one to $Y_d Y_d^\dagger$ which in consequence
enhances the misalignment between $c_Q$ and $Y_d$. We have confirmed
this argument numerically.
In Fig.\ref{5dmfvscan}, we show the dependence
of $\lt| C_4^{\rm 5D-MFV}/C_4^{\rm RS} \rt|$ on the 5D CKM angles
for a reference model with $c_u = (0.68,0.53, -0.06),
c_d = (0.65, 0.6,0.58),c_Q = (0.64, 0.59,0.46)$.
We have
fixed the ratio $r=\beta^L_u/\beta^L_d = 1/4$ and scanned over the
5D CKM angles. We only plot points which result in the correct 4D
CKM matrix and which do not have singular 5D CKM matrices where 5D Kobayashi-Maskawa mechanism would render a CP conserving model (this is
the reason for the lack of points around $\pi/2, \pi, \ldots$). We
clearly see that most of the points give much less suppression than
the naive expectation $r^2 \sim 1/16$. Assuming no tuning in
the 5D CKM matrix we conclude that only the limit $r\rightarrow 0$
seems to sufficiently suppress dangerous FCNCs. This points towards
a symmetry which we explore in the next section.

\section{Complete Alignment via Shining at Leading Order}
\label{sec:models} \setcounter{equation}{0} \setcounter{footnote}{0}

We have seen above that a suppression of the $Y_u Y_u^\dagger$ term
in the expression for $c_Q$ is generically not sufficient to
suppress down-type FCNCs.
It seems that the only simple solution to this problem is to
identify possible symmetry breaking patterns which automatically give
$r_u=0$.
In this case the alignment is complete, irrespective of the
5D mixing angles.
%

Here we consider two different models.
In the first model, we impose SU(3)$_Q\times$SU(3)$_d$
bulk flavor symmetry, while in the second one we impose only the diagonal subgroup
SU(3)$_{Q+d}$.
In both cases, the bulk scalar transforms as the down-type Yukawa
$Y_d$ itself, namely ${\bf 3}_Q\times{\bf \bar 3}_d$
or ${\bf 8}_{Q+d}$ (assuming an adjoint field), respectively.
We do not impose any flavor symmetries for the up sector, so that it is anarchic.
Such scenario can be realized by having a flavor symmetry which is broken
in the bulk by multiple sources but not via flavon fields which transform under the
${\bf 3}_Q\times{\bf \bar 3}_d$ group such as a bulk $Y_u$ field.
The up-type Yukawa coupling is generated only on the IR brane, so that it cannot
feed into the bulk fermion masses $c_{Q,d}$.
In these models the alignment in the bulk between $c_{Q,d}$ and $Y_d$ is complete at leading order.
Higher order effects are discussed in the next section.

\subsection{$SU(3)_Q\times SU(3)_d$ Bulk Symmetry\label{sec:quadraticalignment}}

In the first realistic example we assume that the down sector
transforms under a SU(3)$_Q\times$SU(3)$_d$ bulk flavor symmetry and
that the only breaking is given by $Y_d ({\bf 3}_Q , {\bf \bar 3}_d)$.
Then the expression for the bulk mass parameters
is expected to be of the form
\begin{eqnarray}
c_Q &=& \alpha_Q \cdot \mathbbm{1} + \beta_Q Y_d Y_d^\dagger  \label{cQ}\\
c_d &=& \alpha_d \cdot \mathbbm{1} + \beta_d Y_d^\dagger Y_d \label{cd}\\
m^{(d)}_{ij} &=&\frac{v}{\sqrt{2}} {f_Q}_i {Y_d}_{ij} {f_d}_j.
\label{md}
\end{eqnarray}
Note that in the above we have omitted, for simplicity, higher order terms.
These terms
would not spoil the alignment, but once resummed, they would only lead to a shift in the value of the eigenvalues~\cite{GMFV}.
In the RH up-sector we assume  a generic breaking of the flavor
symmetries by an anarchic Hermitian matrix.

We can find a numerically acceptable solution in the
following way:
\ben
\item
In the basis where $Y_d$ is diagonal, we fix
the input values of the $c_Q$ matrix by requiring that it reproduces
the physical CKM matrix.
In particular, we used three sets of $c_Q$ eigenvalues: the standard choice
\begin{equation}
c_Q=(0.632,0.585,0.430) \,,
\end{equation}
another one with a more composite third generation
\begin{equation}
c_Q=(0.610,0.561,0.200) \,,
\end{equation}
and finally close to fully composite third generation
\begin{equation}
c_Q=(0.602,0.552,0.000) \,.
\end{equation}
\item
Eq. (\ref{cQ}) then fixes $Y_d$ as a
function of $\alpha_Q$ and $\beta_Q$.
\item
Using (\ref{cd}), we obtain an expression for $c_d$ as a function of the four input parameters
$\alpha_{Q,d},\beta_{Q,d}$.
\item
We then scan over these four parameters,
requiring that Eq.(\ref{md}) reproduces the observed SM down-type quark
masses.
\een

Numerical examples resulting in the correct CKM and masses
for the down quarks are given in Table~\ref{tab : adjointexample}.
\begin{table}[htb]
\begin{center}
\renewcommand{\arraystretch}{1.25}
\begin{tabular}{cccccc}
$c_{Q_3}$ & $\alpha_Q$ & $\beta_Q \cdot 10$ & $\alpha_d$ & $\beta_d  \cdot 10^2$ & $Y_d$ \\
\hline
0.000    &  1.258       & -0.239    &   0.774    & -0.228 & (5.237,\,5.433,\, -7.252) \\
0.000    &  -1.186       & 0.318   &   0.528    & 0.324 & (7.498,\,7.392,\, -6.107) \\
0.200 & 0.635 &          -0.733 & 0.629 & -0.263 & (0.591,\,1.009,\,2.437) \\
0.430 & 0.697 & -0.944 & 0.635 & -1.985 & (0.833,\,1.092,\,1.683) \\
0.430 & 0.754 & -0.636 & 0.655 & -1.316 & (1.388,\,1.633,\,2.258) \\
\hline
\end{tabular}\caption{Numerical examples for the model with $SU(3)_Q\times SU(3)_d$ bulk symmetry  resulting in a better than $95\%$ C.L. fit of the mixing angles and masses.}
\label{tab : adjointexample}
\end{center}
\end{table}
\renewcommand{\arraystretch}{1}

We can see that in these solutions the Yukawa couplings are not
 hierarchical, and they are really implementations of the anarchic
 approach with additional bulk alignment. One might be concerned about the
 fact that in these numerical solutions, the correction to the flavor
universal term is not always small (while perturbativity is always
maintained). This would imply that the next correction might
significantly modify the actual numerical values of the
coefficients.
However, as already mentioned, these higher order terms would not spoil the alignment
itself, and therefore the absence of down-type tree-level FCNC is maintained.

\subsection{$SU(3)_{Q+d}$ Bulk Symmetry\label{sec:linearalignment}}
The second model that we are considering employs a diagonal flavor
symmetry $SU(3)_{Q+d}$ in the bulk, which is broken only by an adjoint
spurion $A_d({\bf 8}_{Q+d})$.
The main difference compared with the $SU(3)_Q\times SU(3)_d$ case
is that the alignment between $c_{Q,d}$ and the Yukawa $Y_d$ is not
quadratic but linear, resulting in a very different numerical
solution.
The expressions for the bulk masses consistent with these
symmetries are given by
\begin{eqnarray}
c_Q &=& \alpha_Q \cdot \mathbbm{1} + \beta_Q A_d  \\
c_d &=& \alpha_d \cdot \mathbbm{1} + \beta_d A_d \\
m^{(d)}_{ij} &=&\frac{v}{\sqrt{2}} {f_Q}_i ( \alpha_Y \cdot \mathbbm{1} + \beta_Y  A_d)_{ij}  {f_d}_j
\end{eqnarray}
where $c_u$ is again assumed to be unconstrained. The procedure of
finding numerical solutions for this system is slightly easier:
\ben
\item
Again, we go to a basis where $c_{Q,d}$ and $A_d$ are
diagonal, and fix the same input $c_Q$ as before.
\item
This fully fixes the direction of the adjoint matrix $A_d$ in the space of
$\mathbbm{1},\lambda_3,\lambda_8$,  where $\lambda_{3,8}$ are the diagonal Gell-Mann matrices.
Writing it as $c_Q= a_0 \cdot \mathbbm{1}+a_3 \cdot \lambda_3 +a_8
\cdot \lambda_8$, we find for example that $a_0 = 0.55, a_3=0.033 ,a_8=0.14$ for the
standard choice of $c_Q$.
\item
The expression for $c_d$ can then be written
as $c_d= \alpha_d \cdot \mathbbm{1} +\beta_d \cdot ( \lambda_3
+\frac{a_8}{a_3} \lambda_8)$.
\item
Now we can obtain $f_{c_d}$ as
function of $\alpha_d, \beta_d$, from which one can
calculate the alignment of the $m^d$ mass matrix in the
$\lambda_3,\lambda_8$ space.
\item
Requiring that $m^d$ points in the same
direction in this space fixes the value of $\alpha_d$ for any
input value of $\beta_d$.  This is performed numerically.
\een
A few  simple numerical examples are given in Table~\ref{tab : diagonalexample}. We again find that the perturbative
treatment of Yukawa insertions makes sense and that we do not need
hierarchies in the Yukawa matrices to generate the SM CKM and mass spectrum.

\begin{table}[htb]
\begin{center}
\small
\renewcommand{\arraystretch}{1.25}
\begin{tabular}{ccccccccc}
$c_{Q_3}$ & $\alpha_Q$ & $\beta_Q$ & $\alpha_d$ & $\beta_d\times 10$ & $\alpha_Y$ & $\beta_Y$ & $A_d$ & $Y_d$\\
\hline
0.000    & 0.384      & 1.000    &  0.700    & 1.13 & 7.926 & 0.204 &(0.217,\,0.167,\,-0.384) & (7.984,7.970,7.825)\\
0.000   & 0.384      & 1.000    &  0.685    & 0.998 & 5.930 & 0.957 &(0.217,\,0.167,\,-0.384) & (6.137,\,6.090,\,5.562)\\
0.200    & 0.457      & 0.100    &  0.631    &  0.189 & 1.428 & 0.204 &(1.529,\,1.039,\,-2.568) & (1.740,\,1.640,\,0.904)\\
0.200    & 0.457      & 0.100    &  0.601    & 0.0276 & 0.679 & -0.233 &(1.529,\,1.039,\,-2.568) & (0.323,\,0.437,\,1.278)\\
0.430    & 0.549      & 0.100    &  0.580    & 0.135 & 0.722 & -0.299 &(0.830,\,0.359,\,-1.188) & (0.474,\,0.615,\,1.077)\\
0.430    & 0.549      & 0.100    &  0.590    & 0.195 & 0.823 & -0.281 &(0.830,\,0.359,\,-1.188) & (0.589,\,0.722,\,1.158)\\

\hline
\end{tabular}\caption{Numerical examples for the model with $SU(3)_{Q+d}$ bulk symmetry resulting in a better than $95\%$ C.L. fit of the mixing angles and masses.}
\label{tab : diagonalexample}
\end{center}
\end{table}
\renewcommand{\arraystretch}{1}

\section{Misalignment from loop effects}
 \setcounter{equation}{0} \setcounter{footnote}{0}
\label{sec:loops}

In the above, we have shown that, at leading order, we can completely eliminate the down-type
FCNCs from bulk effects by aligning $c_Q$ and $c_d$ with $Y_d$.
Since we assume that the up-type Yukawa (which
breaks the flavor symmetries needed for the alignment) is an IR
brane field, the alignment in the bulk could be complete. However, on
the IR brane there will be higher dimension operators and loop effects which feed $Y_u$ into $Y_d$,
 so that
the brane induced $Y_d$ is misaligned with the bulk $c_{Q,d}$.
These effects are still
placing very significant bounds on the model, some of which could
actually be comparable to the tree-level bounds in the generic
anarchic model found in~\cite{cfw1}.
The fact that some NLO effects could be very strong is not completely unexpected, since
in order to suppress the tree-level CP violation in the $D$ system~\cite{DCPV} and to obtain a
sufficiently large top mass, one usually needs to push the Yukawas
close to their perturbative bounds, which in turn enhances the
ratio of loop effects and also increases the importance of higher dimensional operators.

First we present the NDA estimates of all the couplings relevant for generating misalignments via loops. Then we present the bounds on the parameters from loop- (and higher dimensional operators-)induced effects:
the first is contributing to the down Yukawa scalars on the IR brane; the second is kinetic mixing for the down-type quarks on the IR brane; the third is kinetic mixing on the UV brane; finally we comment on all other possible sources of additional flavor violation.

\subsection{NDA for everything}

The theory we are considering has an intrinsic cutoff scale $\Lambda$, which implies that the non-renormalizable theory becomes strongly coupled at that scale. In order to have a regime where the theory behaves as a weakly coupled 5D theory, we require that at least the first few KK modes are weakly coupled. We denote the number of weakly coupled KK modes by $\NKK$ and usually require $\NKK\geq 3$, which implies $\Lambda \sim  \NKK/R$ for the unwarped cutoff scale.

\subsubsection*{Brane localized Yukawa coupling}
\begin{figure}
   \begin{center}
      \includegraphics[angle=270,width=4cm]{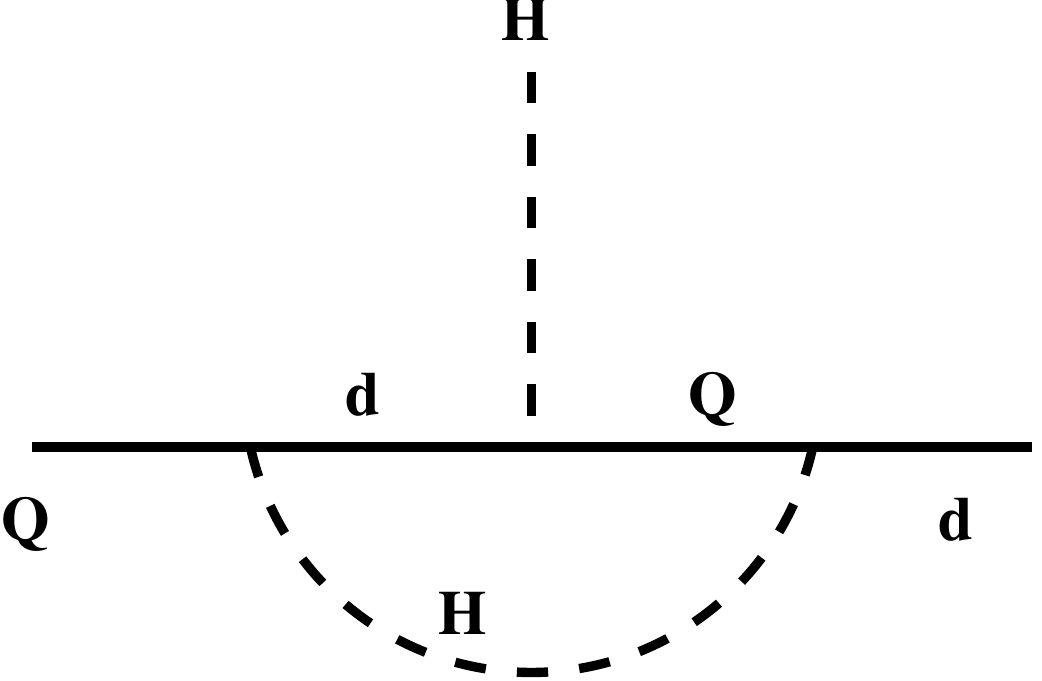}
   \end{center}
   \caption{A one-loop diagram for corrections to a brane Yukawa coupling. \label{fig:NDAbraneyukawa}}
\end{figure}

As a warm-up exercise, let us first consider the NDA bound for a brane Yukawa coupling of the form
\begin{equation}
\int \! d^4x\,\, y_d \bar QH d \bigg{|}_{z=R'}.
\end{equation}
Here $y_d$ is a dimension $-1$ coupling. The NDA bound is obtained by requiring that the one-loop contribution in Fig.~\ref{fig:NDAbraneyukawa} to this coupling is not larger than the tree-level term itself. The estimate for the one-loop correction is given by
\begin{equation}
\Delta y_d^{1-loop} \sim \frac{4 y_d^3}{16 \pi^2} \Lambda^2
\end{equation}
where we have taken into account that the KK modes in the loop are coupled by a factor of $\sqrt{2}$ more strongly than zero modes. The requirement that this does not exceed the tree-level bound will give the upper bound on the dimensionless Yukawa coupling $Y_d = y_d R^{-1}$
\begin{equation}
Y_d< \frac{2\pi}{\NKK} \label{eq:Yd-bound}
\end{equation}
which is the usual bound on the dimensionless Yukawa couplings usually imposed on generic RS flavor models.

\subsubsection*{Brane Yukawas from flavor scalars}

\begin{figure}
   \begin{center}
      \includegraphics[angle=270,width=14cm]{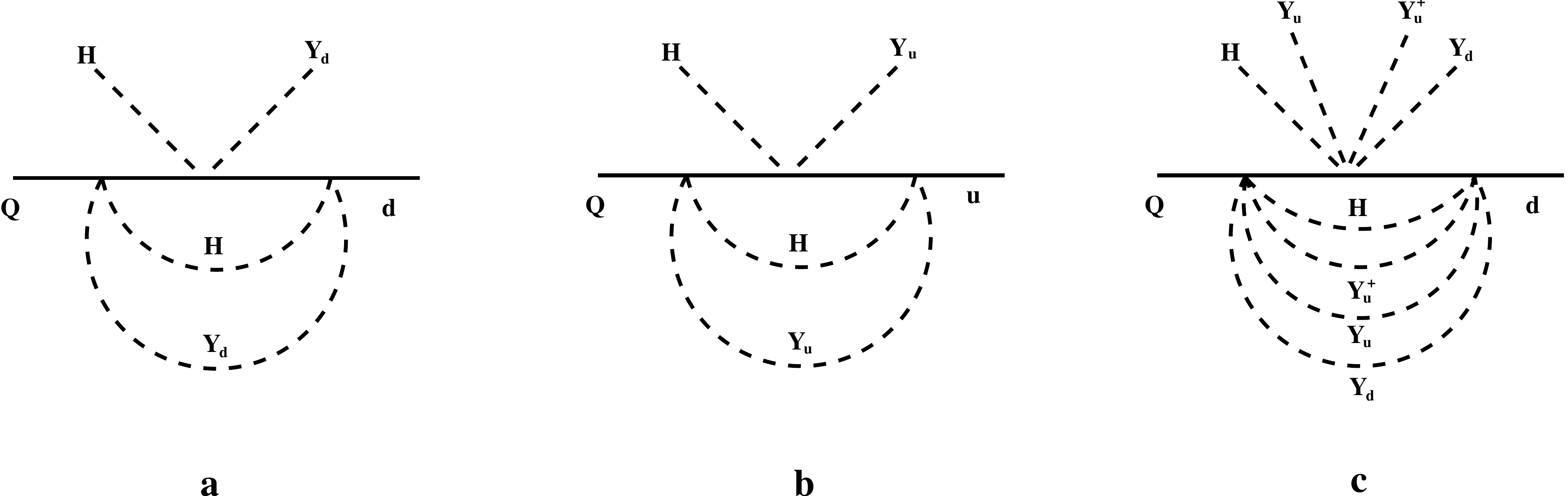}
   \end{center}
   \caption{The leading corrections to the coefficients $\lambda_d$ in {\bf a}, $\lambda_u$ in {\bf b} and $\lambda_d^u$ in {\bf c} used estimating the NDA values of these operators.\label{fig:bulkyukawa}}
\end{figure}

What we are interested in here is the bound on the coefficients of brane localized Yukawa-type operators which originate from insertions of bulk or brane scalar fields breaking the flavor symmetries. As suggested in the previous section, we will assume that the down-type flavor symmetry is broken by a bulk scalar $y_d$, while the up-type by a brane scalar $y_u$. We are then interested in the NDA estimate of the size of the operator involving the $y_u$ field in a down-type Yukawa coupling, which will give rise to misalignment and FCNC's. So we want NDA bounds on the coefficients $\lambda_d,\lambda_u$ and $\lambda_d^u$ for the operators

\beq
S_{IR} \ni \int \! d^4x \left[\lambda_d y_d \bar Q d H+\lambda_u y_u \bar Q u H +
{\lambda^{u}_{d}} \, y_u y_u^\dagger y_d \bar Q d H\right]\bigg{|}_{z=R'}\label{Lin}
\eeq
Since we assume that $y_d$ is a bulk field and $y_u$ is a brane field the coefficients
$\lambda_d,\lambda_u, \lambda^u_d$ are of mass dimension $E^{-5/2,-2,-9/2}$ respectively. The leading corrections (see Fig.~\ref{fig:bulkyukawa}) arise at two loops for $\lambda_d$ and $\lambda_u$, while for
${\lambda^{u}_{d}}$ at four loops. Therefore (using the same rules as before) we find the leading corrections to be:
\begin{equation}
\Delta \lambda_d = \frac{8 \lambda_d^3}{(16 \pi^2)^2} \Lambda^5, \ \ \Delta \lambda_u = \frac{4 \lambda_u^3}{(16\pi^2)^2}\Lambda^4, \ \
\Delta \lambda_d^u = \frac{8 (\lambda_d^u)^3}{(16\pi^2)^4}\Lambda^9.
\end{equation}
The resulting NDA bounds will thus be
\begin{equation}
\lambda_d < \frac{16\pi^2 R^\frac{5}{2}}{\sqrt{8}\NKK^\frac{5}{2}}, \ \ \lambda_u < \frac{16\pi^2 R^2}{2 \NKK^2}, \ \
\lambda_d^u< \frac{(16\pi^2)^2 R^\frac{9}{2}}{\sqrt{8} \NKK^\frac{9}{2}} .
\label{eq:NDAbounds}
\end{equation}
As a consistency check, one can switch off the value of $\lambda^u_d$ and regenerate it at one loop via the $\lambda_{d,u}$
couplings. The contribution is generated at one-loop and is proportional to $4 \lambda_u^2 \lambda_d \NKK^2/(R^2 16\pi^2)$.
Substituting the NDA values of $\lambda_{u,d}$ in this expression, we indeed find the NDA value of $\lambda_d^u$.

\subsubsection*{Scalar VEVs}

In order to find the NDA value of the effective dimensionless Yukawa couplings $Y_{u,d}$ in our model, we have to find (besides the coefficients $\lambda_{u,d}$) also the NDA values of the bulk and brane scalar VEV's $\langle y_{u,d}\rangle$. For this we assume that there is a brane mass of order the cutoff scale generated, and in addition a brane localized quartic self-interaction. For the down-type bulk scalar $y_d$
\begin{equation}
V(y_d)_{IR}=\Lambda y_d^\dagger y_d +\lambda_4 |y_d|^4
\,,
\end{equation}
where $\lambda_4$ is of mass  dimension $-2$. The NDA value of $\lambda_4$ is obtained in a similar way to the above derivations: there is a one loop quadratically divergent correction
\beq
\Delta \lambda_4=4 \lambda_4^2 \frac{\Lambda^2}{16\pi^2}
\eeq
Hence the NDA value of $\lambda_4$ is
\beq
\lambda_4\sim \frac{4 \pi^2 R^2}{\NKK^2} .
\eeq
Thus the NDA value of the scalar VEV $\langle y_d \rangle$ is
\beq
\langle y_d\rangle \sim \frac{\Lambda^\frac{1}{2}}{\lambda_4^\frac{1}{2}} = {\NKK^{3/2}\over 2\pi R^{3/2}}\,.
\eeq
We also note that since $y_u$ is an IR field, it follows the ordinary 4D power counting,
and hence
\beq
\langle y_u\rangle \sim {\NKK\over 4\pi R}\,.
\eeq

We have two important remarks regarding the NDA values of the VEVs:
\begin{itemize}
\item While all previous NDA estimates were {\it upper bounds} requiring that couplings don't blow up before the cutoff scale, finding the NDA value of the VEV $\langle y_{d,u}\rangle$ involves minimizing a potential assuming the maximal NDA size quartic coupling. The VEV would actually {\it increase} with a decreasing quartic self-coupling. Thus for VEVs the NDA size is {\it not} an upper bound, merely the natural value. If one allows some cancellation between the tree-level and loop induced quartic, then the VEV can be {\it increased} without leaving the region of validity of the effective theory, but at the price of a tuning. For example, if there is a 50\% cancelation in $\lambda_4$ the VEV can be increased by a factor of $\sqrt{2}$.

\item One does expect the size of the VEV to be set by the local infrared potential, even if the VEV started out with a different size on the UV brane. The reason is that one usually finds that by solving the bulk equations for a scalar VEV one usually finds that the VEV will "twist" between the UV and the IR brane values, ie. interpolate between the two natural NDA values set on the branes.

\end{itemize}

\subsubsection*{Effective Yukawas from scalar VEVs}

The NDA value effective Yukawa coupling for the bulk scalar is
\begin{equation}
Y_d = \lambda_d \langle y_d \rangle R^{-1} = \frac{2 \sqrt{2} \pi}{\NKK} .
\end{equation}
This is a factor of $\sqrt{2}$ larger than the NDA bound for a brane localized Yukawa, Eq.~(\ref{eq:Yd-bound}), whose origin can be traced back to the fact that in the NDA bound for $\lambda_d$ there is one brane localized field running in the loop which does not pick up an enhanced coupling. In addition, we have explained above that raising the VEV $\langle y_d \rangle$ does not imply leaving the domain of validity of the effective theory, but rather a certain amount of tuning between the tree-level and loop induced quartics for $y_d$ on the IR brane. For example, to reach $Y_d =10$ with $\NKK=3$ one needs a tuning of about 10\%.

Similarly, if $Y_u$ is generated from a brane scalar as proposed before, its NDA value is
\begin{equation}
Y_u = \lambda_u \langle y_u\rangle R^{-1} = \frac{2\pi}{\NKK}.
\end{equation}

\subsection{Loop induced  corrections to the down Yukawa coupling}

\begin{figure}

   \begin{center}
      \includegraphics[angle=270,width=4cm]{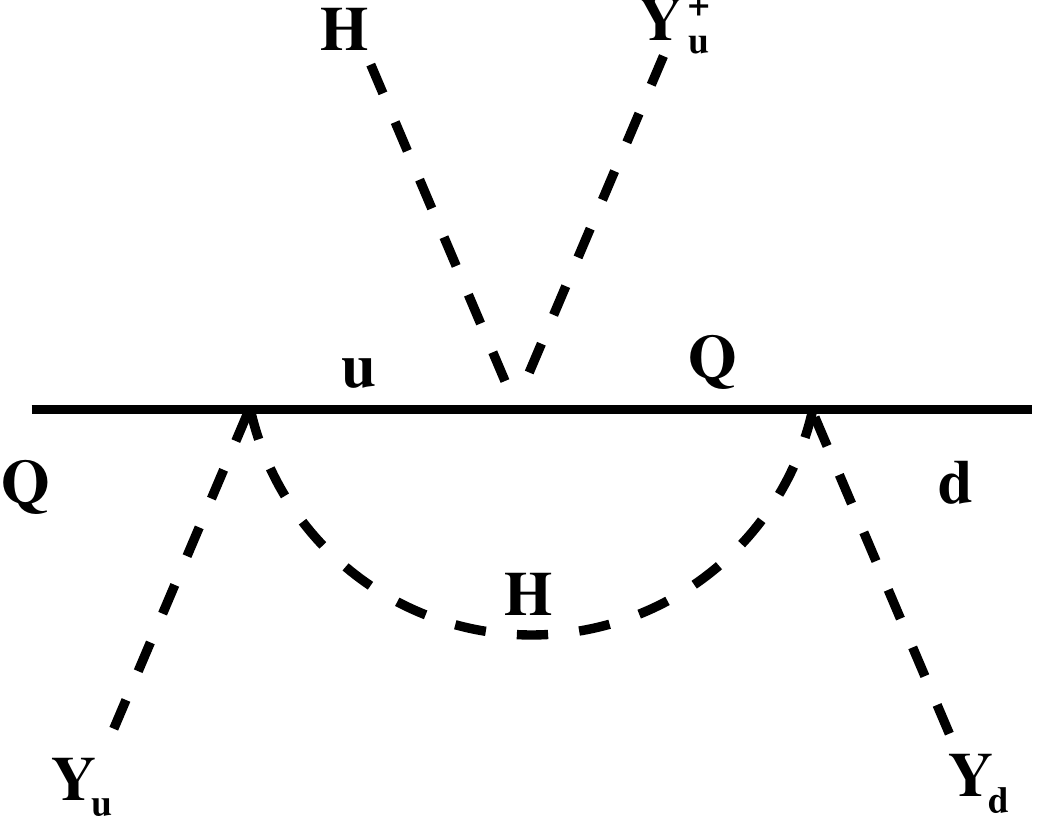}
   \end{center}
   \caption{A 1-loop diagram on the IR brane responsible for the misalignment between the bulk and the brane.\label{fig:badguy}}
\end{figure}

We are now able to estimate the misaligning effects of the IR brane loop corrections (or similarly the ones from higher dimensional operators). We have already estimated the NDA size of the $\lambda_d^u$ operators in Eq.(\ref{eq:NDAbounds}). This teaches us that the effective down-type Yukawa $Y_d$
is shifted to:
\begin{equation}
Y_d\to Y_d^{(tot)}=Y_d+ \lambda_d^u \langle y_u\rangle \langle y_u^\dagger \rangle \langle y_d \rangle R^{-1} =Y_d+
\left( \frac{\lambda_d^u}{\lambda_d\lambda_u^2}\right) Y_u Y_u^\dagger Y_d R^2.
\end{equation}
Using the NDA values of the $\lambda$'s, we find
\begin{equation}
Y_d^{(tot)} \sim Y_d \left(1+ \frac{\NKK^2}{4\pi^2} Y_u Y_u^\dagger\right),
\label{suppression}
\end{equation}
which is generated by the 1-loop diagram in Fig.~\ref{fig:badguy}. So the suppression parameter for the misaligning effects is given by
\begin{equation}
\epsilon_u =  {Y_u^2 \NKK^2\over 4 \pi^2}
\end{equation}
where ${Y}_u$ is an average value of the elements of the $Y_u$ matrix. Note that this diagram is generically present in any alignment model where the Yukawas are dynamical fields, and oparators of this sort cannot be forbidden by any symmetry. Since our theory is just on the verge of being perturbative, such corrections can be very dangerous. In fact, a recent analysis~\cite{DCPV} shows that bounds on CPV in $D-\bar{D}$ mixing requires $Y_u > 1.6$, and obtaining a sufficiently heavy top mass would also suggest a large value of $Y_u$ ($Y_u=0.5$ is the lowest possible value when both the LH and RH top fields are fully composite). This implies that the suppression parameter $\epsilon_u$ is numerically not that small. For $Y_u=1.6$ and $\NKK=3$ we find $\epsilon_u =0.58$ (while for $\NKK=2$ we would get $\epsilon_u=0.26$).

Due to the non-trivial flavor structure of (\ref{suppression}) one needs to analyze carefully the down-type mass matrix in order to actually read off what the suppression of the flavor violating KK gluon couplings are.
An easy way to do this is to evaluate the misalignment (in the basis where both the wave functions $f_{Q,u,d}$ and $Y_d$ are diagonal) from the left:
\begin{eqnarray} A_Q&\equiv& f_Q
Y_d^{(\rm tot)} f_d^2 \left(Y_d^{(\rm tot)}\right)^\dagger f_Q\nonumber\\
 &\sim&
f_Q Y_d f_d^2 Y_d f_Q + {4 \NKK^2\over 16 \pi^2} \,f_Q  \left[  Y_d
f_d^2 Y_d {Y}_u Y_u^\dagger + Y_u  Y_u^\dagger
Y_d f_d^2 Y_d\right] f_Q\nonumber \\
&+&  \left({4 \NKK^2\over 16 \pi^2}\right)^2 \,f_Q Y_u  Y_u^\dagger
Y_d f_d^2 Y_d Y_u Y_u^\dagger f_Q\label{HiggsQ},
\end{eqnarray}
and from the right:
\begin{eqnarray} A_d &\equiv& f_d \left(
Y_d^{(\rm tot)}\right)^\dagger f_Q^2  Y_d^{(\rm tot)} f_d \nonumber\\
&\sim&  f_d
Y_d f_Q^2 Y_d f_d+ {4 \NKK^2\over 16 \pi^2} \,f_d  \left[  Y_d f_Q^2
Y_u  Y_u^\dagger Y_d +  Y_d Y_u Y_u^\dagger f_Q^2 Y_d\right] f_d
\nonumber \\
&+&  \left({4 \NKK^2\over 16 \pi^2}\right)^2 \,f_d   Y_d Y_u
Y_u^\dagger f_Q^2 Y_u Y_u^\dagger  Y_d f_d \label{Higgsd},
\end{eqnarray}

The KK gluon coupling is diagonal but not universal in the original gauge eigenstate basis and is proportional to
$g_{s*} f_{q_i}^2$ for the $q_i$ quark.
The FCNC contributions are then proportional to the amount of rotation
we need to perform in order to diagonalize (\ref{HiggsQ}-\ref{Higgsd}).
The terms linear in the loop factor have the same wave-function and
RS-GIM structure as the tree-level terms in the generic RS model
without alignment, so the only difference would be the suppression
factor $\epu$. So the off-diagonal coupling of the LH down-type fields will be given by
\begin{equation}
g_{12}^L=g_{s*} \epsilon_u f_{Q_1} f_{Q_2} .
\end{equation}
When considering a similar effect in the right handed fields, one also has to take into
account another important fact: the flavor changing effects in the terms
proportional to the square of the loop factor can also go through
the third generation, and those have an additional enhancement
factor of $(f_{d_3}/f_{d_2})^2$ (for the LH fields) and $(f_{Q_3}/f_{Q_2})^2$ (for the RH fields).
Expressing
these in terms of the physical quark masses and the Cabbibo angle, we
find that for $A_d$, the term with the two-loop factor is actually
larger than the term with the one loop-factor (while for $A_Q$ they
are roughly comparable). In fact since $f_{Q_3}/f_{Q_2} \sim 1/\sin^2\theta_c =25$ this will be the leading term in the mass matrix, which implies that for the RH couplings one reverts back to the anarchic case - the misalignment in the RH sector is complete. So the RH couplings of the KK gluon will be unsuppressed compared to the standard RS GIM result
\begin{equation}
g_{12}^R=g_{s*} f_{d_1}f_{d_2}.
\end{equation}
The effect of the KK gluon exchange is then $g_{12}^L g_{12}^R/M_G^2$, where as usual one converts to physical masses and mixing angles using the expressions of anarchic RS flavor (which are also applicable here), giving rise to
\begin{equation}
C_{4K}\sim \epsilon_u \frac{g_{s*}^2}{M_G^2} \frac{1}{Y_d^2}\frac{2 m_d m_s}{v^2}
\label{C4aligned}
\end{equation}
Thus naively one does not gain much (just one factor of $\epsilon_u$) by going to alignment models. This is however not quite the case. We argue below that there are two ways using the alignment structure to further reduce the size of these corrections.

\subsubsection*{Suppressing $\epsilon_K$ via enhanced $\langle y_d \rangle$}

In order to numerically compare the bound on (\ref{C4aligned}) to the most stringent bound on simple anarchic RS flavor from \cite{cfw1}, we also need to take into account the fact that here the $Y_d$ coupling can be taken to be somewhat larger than the NDA bound in anarchic RS, since it is originating from a VEV of a bulk scalar. In this case, as we saw, raising the effctive $Y_d$ above the NDA value does not imply leaving the regime of validity of the effective theory, but rather some amount of tuning in the scalar potential setting the IR brane VEV.
In the calculation of the bound on the KK gluon mass in~\cite{cfw1}, the $Y_d$ coupling was fixed to $Y_d \sim 3$, and we have seen that $C_{4K}$ in Eq.(\ref{C4aligned}) scales as $1/Y_d^2$. The bound on the KK scale from this KK gluon mediated LLRR operator is about $10$~TeV, if the running of the UV brane induced localized gluon kinetic term is taken into account (while it is about 20 TeV without that effect).
If we want to reduce that bound to about 3 TeV (the lowest allowed by generic electroweak precision
constraints), then one would need to raise $Y_d$ to about
\begin{equation}
Y_d \gtrsim 3 \left( \frac{ 10 \ {\rm TeV}}{ 3 \ {\rm TeV}}\right) \sqrt{\epsilon_u} \sim 7.6 \left( \frac{\NKK}{3} \right).
\end{equation}
Thus, raising the effective $Y_d$ to about $Y_d \geq 7.5$ (for $\NKK=3$) is sufficient (for $\NKK=2$, one needs
$Y_d \geq 5.1$). This would imply that the $\langle y_d \rangle$ VEV is about 2.5 times its NDA value, corresponding to a tuning of about 15\% in the IR brane localized quartic. For $\NKK=2$, $\langle y_d\rangle$ is just 1.1 times the NDA value, and the fine tuning is almost completely absent (just 75\%). Note, that numerical results for aligned solutions with sufficiently large $Y_d$ can be found both for the quadratic alignment (section \ref{sec:quadraticalignment}) and for the linear alignment model (section \ref{sec:linearalignment}).
Also note that contributions to dipole operators from higher dimensional operators
(such as EDMs, $b\to s\gamma$ and $\epsilon'/\epsilon_K$, see~\cite{GIP} for a recent discussion) are suppressed by $\epsilon_u$ and even further suppressed
in the two-higgs doublet model described in the following section, and hence are probably consistent with the current bounds.

\subsubsection*{Two-Higgs Doublet Model}
\label{2HDM}
\begin{figure}
     \begin{center}
      \includegraphics[angle=270,width=8cm]{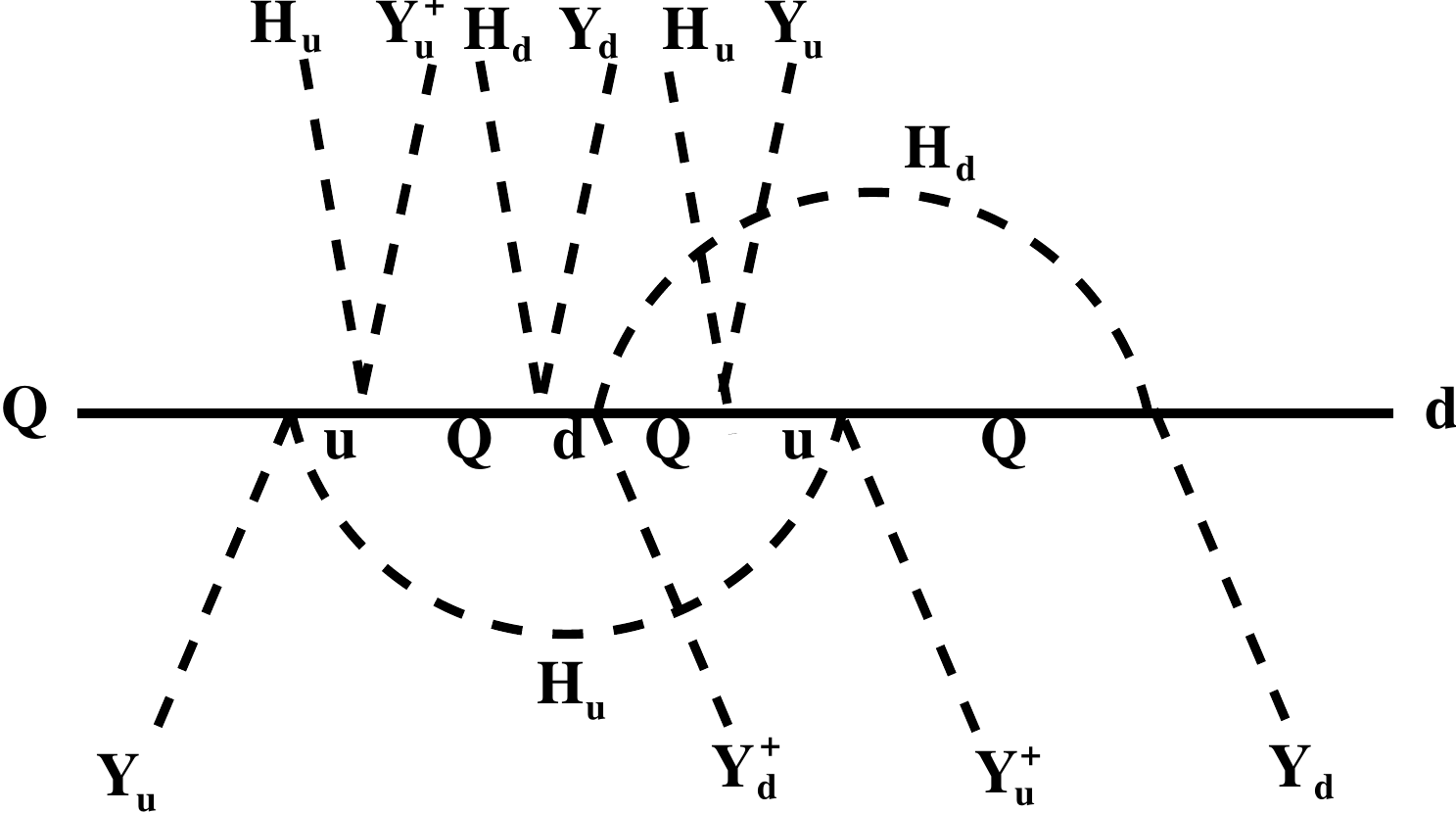}
   \end{center}
   \caption{A 2-loop quartically divergent diagram contributing to the misalignment even in the 2HDM. By closing up additional Higgs and Yukawa scalar lines one can obtain higher loop (but more divergent) contributions.\label{fig:2loop2HDM}}
\end{figure}

Above we saw that the generic one-loop correction to the down Yukawa is quite large, so the final suppression is only proportional to $\epsilon_u$ (and possibly further suppressed via enhanced down Yukawa couplings). One may try to suppress the misaligning terms further by introducing more structure. However, it is clear from the outset that there is no symmetry that can forbid the operator $\bar Q Y_u^\dagger Y_u Y_d d H$ if the $\bar Q Y_d d H$ is allowed, so this program can only postpone the issue to higher loops, which numerically might still be sufficient. One possible way to realize
this goal is if there were separate up and down type IR-localized Higgses: the $H_u$ coupling to $\bar QY_u u$ and the $H_d$ to $\bar QY_d d$. Then the dangerous one-loop diagram would be absent. However, in the limit when there is no coupling between $H_u$ and $H_d$ there is also a  new U(1)$_{PQ}$ global symmetry which would give rise to a weak-scale  axion. As usual this symmetry has to be broken, but the breaking scale (for example via a $\mu^2 H_u H_d$ term) can be much lower than the KK scale $1/R'$, which will imply that the one-loop diagrams will be suppressed by $(\mu /\mKK)^2$. If $\mu \sim 100$ GeV, then we gain at least around $1/100$ suppression and the dangerous one-loop diagram is rendered harmless.
However, since there is no actual symmetry, the dangerous operator will indeed be generated at higher loops. For example, at 2 loops the diagram in Fig.~\ref{fig:2loop2HDM} will be quartically divergent, and one can close up legs to get higher loop diagrams with higher power divergences generated. While the 2-loop diagram will have additional Higgs VEVs suppressing the contribution, one can close up those Higgs legs and obtain terms which have again $\epsilon_u = \frac{Y_u^2\NKK^2}{4\pi^2}$ as the expansion coefficient. While we have not attempted to study systematically the higher loop induced misaligning corrections, we can gain some idea of their magnitudes by estimating the magnitudes of the misaligning effects obtained by closing up legs on the diagram in Fig.~\ref{fig:2loop2HDM}. We find the following suppression factors in $C_{4K}$ (for
$Y_d=3$, $Y_u=1.6$, $\mKK=3$~TeV, $\mu=100$~GeV,
taking into account that the Higgs VEVs in a 2HDM are smaller, and choosing the best case $\tan \beta =1$ for the numerical results):

\[
\begin{array}{|c|c|c|c|} \hline
\# {\rm of \ loops} & {\rm parametric \ suppression} & \NKK=3  & \NKK=2 \\ \hline
1-{\rm loop} & \epsilon_u \left( \frac{\mu}{\mKK}\right)^2 & 6 \cdot 10^{-4} & 3 \cdot 10^{-4} \\
2-{\rm loop} & \frac{1}{\cos^2\beta} (2\epsilon_u \frac{v}{\mKK} Y_d)^2  & 0.16 & 0.03 \\
3-{\rm loop} & \frac{1}{\cos^2\beta} \epsilon_u^3 \frac{Y_d^2}{Y_u^2} & 1.4 & 0.12 \\
4-{\rm loop} & \frac{1}{\cos^2\beta} \epsilon_u^4 \frac{Y_d^2}{Y_u^4} & 0.31 & 0.01 \\
5-{\rm loop} & \frac{1}{\cos^2\beta} \epsilon_u^5 \frac{Y_d^2}{Y_u^6} & 0.07 & 10^{-3} \\ \hline
\end{array}
\]
We can see that the suppression is sufficient in all terms investigated if $\NKK=2$, however for $\NKK=3$ the 3-loop contribution to the misaligning effects leading to $C_{4K}$ is actually unsuppressed compared to the RS-GIM case.

\subsection{Loop Induced Kinetic Mixings}

\begin{figure}

   \begin{center}
      \includegraphics[angle=270,width=12cm]{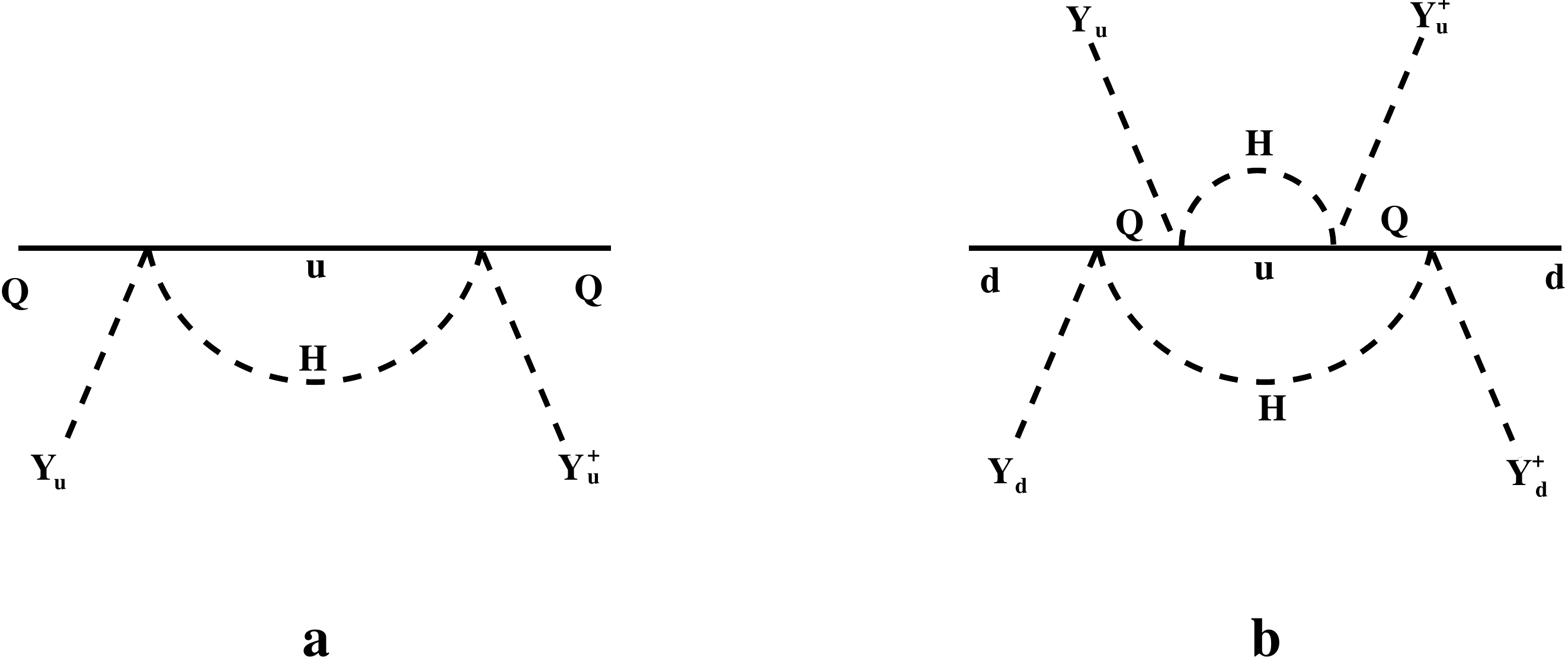}
   \end{center}
   \caption{Corrections to the brane kinetic terms that will result in misaligned couplings, at one loop for $Q$ in a and at two loops for $d$ in b.\label{fig:BKTcorr}}
\end{figure}

The second potentially dangerous brane localized loops contribute to
kinetic mixing terms for the left-handed ($Q$) fields (at one loop) and right-handed
fields (at two loops).   The relevant diagrams are shown in Fig.~\ref{fig:BKTcorr}.

The NDA estimate for the one-loop induced left-handed
kinetic mixing term is
\begin{equation}
 \left(\bar Q D
\hspace*{-.25cm}\slash Q\right)^{(\rm loop)}_{IR}\sim \bar Q D
\hspace*{-.25cm}  \slash \, {2 \NKK \over 16 \pi^2} Y_u Y_u^\dagger\,
Q\,. \end{equation}
For the RH  down quarks the NDA estimate for
the two-loop contribution is
\begin{equation} \left(\bar d D
\hspace*{-.25cm}\slash d\right)^{(\rm 2-loop)}_{IR}\sim \bar d D
\hspace*{-.25cm}  \slash \, \frac{4 \NKK^3}{ (16 \pi^2)^2} Y_d Y_u
Y_u^\dagger  Y_d \, d\,. \end{equation} The off-diagonal kinetic
terms will directly contribute to the off-diagonal KK gluon
coupling, since the kinetic term contains a covariant derivative
including the KK gluon on the brane. Therefore the coefficient of the LR
operator contributing to $\epsilon_K$ (compared to generic magnitude for the anarchic case)
is suppressed by the factor
\begin{equation}
 g_{LLRR}/g_{RS1}\sim  \frac{8 \NKK^4}{(16 \pi^2)^3} Y_d^2   Y_u^4 = \frac{\epsilon_u^2}{2} \frac{Y_d^2}{16\pi^2}
\end{equation}
which for $\NKK=3$ is already a suppression by 1/35, thus putting it safely within the bounds.

\subsection{UV Brane Kinetic Terms}

Next we consider the effects of flavor violation on the
UV brane. The analysis here is quite general, and does not rely on
the specific alignment structure of the bulk mass parameters. We
will allow arbitrary breaking of the flavor symmetries, and assume
that the main effect of this is to induce a kinetic mixing matrix
for the quarks on the UV brane. The only assumption we are making is
that the localized kinetic terms do not dominate over the bulk
kinetic terms, but could be an $\mathcal{O}(1)$ correction
(otherwise they would significantly change the mass hierarchy that
is assumed to come from the wave function overlaps). This is
equivalent to the assumption that the elementary/composite ratio of
the SM fermions is determined by the bulk $c$'s, which a priori can
be modified by large localized kinetic terms (which is the approach
taken in~\cite{xdgim}).
Therefore, we write
\begin{equation}
\mathcal{L}_{UV}= K_{ij}  R \bar{\psi}_L^j \gamma_\mu D^\mu \psi_L^i
\end{equation}
where $K$ is a dimensionless kinetic mixing matrix. The kinetic
terms of the zero modes can be diagonalized via a hermitian rotation
$H=U N^{-\frac{1}{2}} U^\dagger$, such that
\begin{equation}
1= H(1+\tilde{f} K \tilde{f} )H,
\end{equation}
where the $\tilde{f}$'s are the dimensionless fermion wave functions
on the UV brane
\begin{equation}
\psi^i(z=R)=\frac{1}{\sqrt{R}} \tilde{f}^i.
\end{equation}
We can express $H$ as
\begin{equation}
H=\left( 1+ \tilde{f}K\tilde{f}\right)^{-\frac{1}{2}}.
\end{equation}
Now we need to rotate the KK gluon couplings using $H$. Using the
approximate expression for the overlap integral of the KK gluon with
the bulk fermions, and adding the contributions from the UV induced
kinetic term (which also contains a coupling to the gluon) and using
the fact that the KK gluon wave-function on the UV brane is
$1/\log(R'/R)$, we obtain
\beqa
g_{ffG} &\approx& g_{s*} H \left[ -\frac{1}{\log (R'/R)} + f_c^2 \gamma
(c) -\frac{1}{\log (R'/R)} \tilde{f}K\tilde{f}\right] H\nonumber\\
&=& g_{s*}
\left[ -\frac{1}{\log (R'/R)} +H f_c^2 \gamma (c) H \right],
\eeqa
We can see that there is an automatic RS-GIM suppression for the UV
effects of the UV brane kinetic terms too! This is not so
surprising: both the KK gluon and the SM fermions can be thought of
as mixtures of elementary and composite fields. The elementary KK
gluon is the KK gluon at the UV brane, which must have flavor
universal couplings. So all the effects of the kinetic terms must go
through the composite part of the KK gluon (and the SM fermions),
which will necessarily imply that the RS-GIM protection, which is
what we have explicitly found here. We can go a step further and
estimate bounds on the off-diagonal terms of $K$, assuming it is a
subdominant effect compared to 1. In that case
\begin{equation}
H\approx 1- \frac{1}{2} \tilde{f} K \tilde{f} .
\end{equation}
The ratio of the LR operator $C_4^K$ compared to the standard
anarchic case will be
\begin{equation}
 g_{LLRR}/g_{RS1}\sim   \frac{1}{4} K_{12}^L K_{12}^R
 \frac{m_s}{m_d} .
\end{equation}
Again requiring that a 3 TeV KK gluon is satisfying the bound gives
the constraint
\begin{equation}\label{Kbound}
K_{ij} \lesssim 0.1.
\end{equation}
The relative contribution of third generation
effects at each of the vertices is given by
\be
 \frac{ K_{13} K_{32}}{2 K_{12} }\tilde{f_{c_3}}\!{}^2  \frac{f_{c_3}^2}{f_{c_2}^2} =  \frac{(1-2c_3) (R/R^\prime)^{1-2c_3} }{1-(R/R^\prime)^{1-2c_3}}  \begin{cases}
 \frac{ K_{13}^L K_{32}^L}{2 K_{12}^L } {1 \over\lambda^4}\\ \\
   \frac{ K_{13}^R K_{32}^R}{2 K_{12}^R } ({m_b \lambda^2 \over m_s})^2
\end{cases}
\ee
we see that third generation effects are exponentially suppressed for $c_3 < 1/2$. Even for $c_3 > 1/2$,
the extra power of $K$ renders third generation contributions subdominant (we have assumed that the entries of K satisfy the bound in (\ref{Kbound})).
We learn that if $K$ were loop generated, then the extra loop factor would make
these bounds satisfied easily.
Of course generically there is no
reason for the $K$ to be loop suppressed. In the alignment model, one
can of course eliminate all the FCNC's from the UV brane by
postulating that the only source breaking the flavor symmetries is
again proportional to the $Y_d$ field itself. In this case, the
boundary kinetic term is itself aligned with the bulk $c_{Q,d}$
and the Yukawa coupling, and no flavor violation would occur.

\subsection{More Contributions to FCNC}

In our setup we have at least two additional sets of fields that
could potentially lead to flavor violating effects:

- the bulk scalar $y_d$ which is responsible for the alignment

- bulk flavor gauge bosons (which are the gauge fields for the bulk
flavor symmetries SU(3)$_Q\times$SU(3)$_d$).

\begin{figure}

   \begin{center}
      \includegraphics[angle=270,width=12cm]{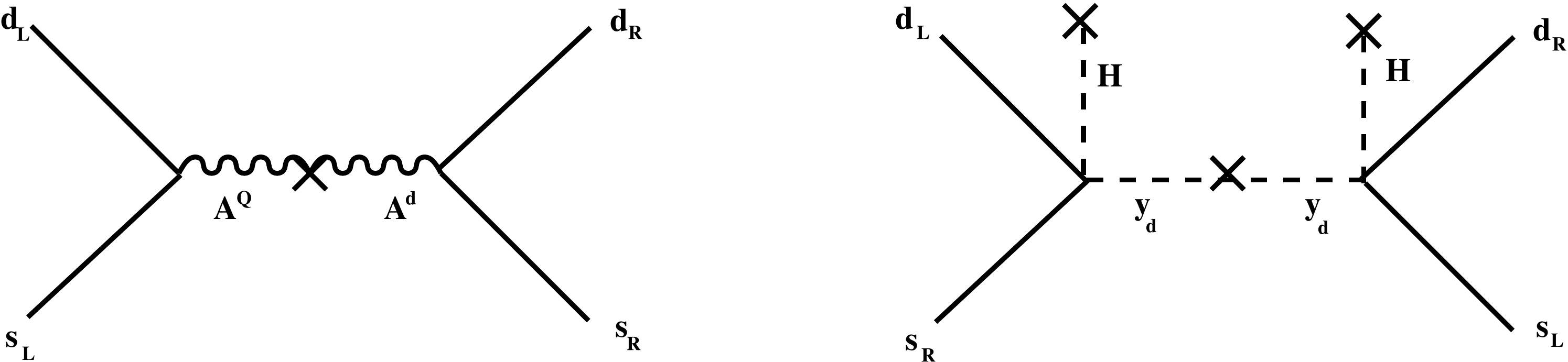}
   \end{center}
   \caption{Diagrams contributing to LR down-type FCNCs.
   These interactions are induced by exchanging either flavor gauge bosons (left) or flavor
   scalars (right). The insertions are flavor violating mass terms, which in the alignment models are proportional
   to $Y_d$.\label{fig:moreFCNC}}
\end{figure}

The couplings of these fields (before $y_d$ obtains a VEV) are
flavor invariant, since the flavor symmetries are assumed to be
gauge symmetries. Thus the only source of flavor violation can be
due to the VEV of the scalar field $y_d$ which is generating flavor
violating masses for the flavor gauge bosons, the down Yukawa
coupling, and potentially also flavor violating masses for the
dynamical scalar KK modes. Exchange of the heavy (TeV-scale) gauge
and scalar modes can in principle give rise to the most dangerous
flavor changing LR four-Fermi operators in the down sector, {\eg}
from diagrams of the form presented in Fig.~\ref{fig:moreFCNC}.
However, for the alignment models presented in this paper the only
source of flavor violation in the bulk is due to the VEV $\langle
y_d\rangle \sim Y_d$, so the alignment extends to all the effects in
these sectors too: the flavor gauge bosons and the KK modes of $y_d$
will be flavor diagonal in the basis where the down-quark masses are
diagonalized, thus no additional FCNC's will be introduced. In the
general case with multiple bulk scalars $\propto Y_u,Y_d$ there will
be (RS-GIM suppressed) effects comparable to the size of the
KK-gluon exchange generated (depending of course on the size of the
5D gauge flavor gauge coupling). We leave for future work the detailed study of the
properties of these fields and their flavor violating effects in the
general case without alignment.

%

\section{Conclusions}
\label{sec:conclusion} \setcounter{equation}{0}
\setcounter{footnote}{0}

In this work, we have shown that warped extra dimensions may provide the solution both to the hierarchy problem and the flavor puzzle, while being accessible at the LHC.
This can be achieved by invoking flavor alignment between the down-type 5D Yukawas and the bulk fermion masses.
Such alignment reduces the flavor problems which are usually present in low-scale models of new physics.
In order to have 5D flavor alignment in the bulk, we assume a flavor breaking source at the UV brane,
and promote the Yukawa couplings to bulk scalar fields, so that they can shine the flavor violation down
the bulk.
We have discussed two models with alignment which are improved versions of the generic 5DMFV model.
Potential problems of this idea - namely loop induced misalignment, flavor violation from UV brane kinetic terms, and flavor violation induced by the new bosons - were shown not to spoil the alignment.
Our framework is fairly predictive in its flavor structure which should allow to test it, possibly even in the near future.
One of the most robust prediction is that the up sector is anarchical.
This should induce CPV in the $D$ system which is just around the corner.
Furthermore, top FCNC contribution via $t_R\to c Z$ is around the LHC reach~\cite{Agashe:2006wa}.
However,
in models where $t_R$ transforms trivially under the custodial symmetry~\cite{newcustodial} to protect $Z\to b\bar b$, this effect is
absent. It is interesting that the same selection of trivial RH up type quark representation also eliminates (to leading order)
Higgs mediated FCNCs~\cite{HFCNC} which correlates the smallness of the RS contributions to $\epsilon_K$ with
the absence of top flavor violation.
In addition we note that the alignment is more effective in suppressing flavor violation between the third generation and the light ones
(the enhancement which is proportional to the ratio of the third to second fermion compositeness fraction is absent).
Thus, our model actually predicts that these transitions are suppressed by $\epsilon_u\sim 0.5$ compared to the anarchic case.
It suggests a discovery of CP violation in $D^0-\bar D^0$ mixing and a small deviation from the SM predictions
in the the down sector might provide a way to verify or exclude our scenario.

\section*{Acknowledgments}

We thank Yuval Grossman for collaboration at early stages of this work.
We also thank Kaustubh Agashe  for useful discussions
and comments.  The research of C.C. has been supported in part by the NSF grant PHY-0757868, and in part by a U.S.-Israeli BSF grant. G.P. is supported in part by the
Peter and Patricia Gruber Award.
Z.S. is supported in part by the US Department of Energy under contract DE-FG03-97ER40546.

\appendix

\section{Fermion zero mode with a bulk mass from shining}
We show how the wave function of the fermion zero mode is affected
by contributions from a bulk scalar field. The action for a bulk
scalar field $\phi$ with a Yukawa v.e.v. on the UV brane is
\beqa
   S &=& \int d^5x\,\lt(\frac{R}{z}\rt)^5\lt\{\lt(\frac{z}{R}\rt)^2\half
   \lt[\partial_\mu\phi\partial^\mu\phi-(\partial_z\phi)^2\rt]-\half m^2\phi^2\rt\}
   \nonumber\\
   &-& \int d^4x\,\lt(\frac{R}{z}\rt)^4\, \frac{\lambda}{2 R} \, (\phi - Y)^2 \Big\vert_{R}
\eeqa With the boundary conditions \beqa
   0= \partial_z\phi \Big\vert_{R'}, \qquad
    0 = \lt( \partial_z\phi - \frac{\lambda}{R} \, (\phi - Y) \rt)\Big\vert_{R},
\eeqa we get \beqa
   \phi(z) = Y \lt(\frac{R}{z}\rt)^\epsilon \, .
\eeqa where we have set $\epsilon = \sqrt{(m R)^2 + 4} - 2$ and
assumed $\epsilon \ll 1 \ll \lambda$.
The bulk mass term of the fermion receives additional contributions
from shining \beqa
 \int d^5 x
\left(\frac{R}{z}\right)^5  M  \psi \chi & \rightarrow &
\int d^5 x \left(\frac{R}{z}\right)^5  \psi \chi \lt(M + \alpha \frac{\phi^\dagger \phi}{\Lambda^2}+\ldots\rt) \\
& \rightarrow & \int d^5 x \left(\frac{R}{z}\right)^5 \psi \chi
\lt(M + \alpha \frac{Y^\dagger Y}{\Lambda^2} \lt(\frac{R}{z}\rt)^{2
\epsilon}+\ldots\rt) \eeqa Setting $
  \beta =\alpha \frac{Y^\dagger Y}{ R \,\Lambda^2} ,
$ the EOM for the left-handed zero mode becomes \beqa
   \lt(\partial_z - \frac{2-c   }{z}\rt) f_{L}(z) + \frac{1}{z} \beta \lt(\frac{z}{R}\rt)^{-2 \epsilon}  f_{L}(z) =0,
\eeqa where we have set $c= M R$. The solution is given by \beqa
f_{L}(z) = N(\epsilon) \,z^{2-c} \exp \lt[ \frac{\beta} {2 \epsilon
}\lt(\frac{R}{z}\rt)^{2 \epsilon } \rt] &=& \tilde N(\epsilon)\,
z^{2-c-\beta} \lt( 1 + \beta \epsilon \lt(  \ln\frac{R}{z}\rt)^2
+\ldots  \rt) \eeqa
\begin{figure}[tb]
\begin{center}
 \includegraphics[width=7.5cm]{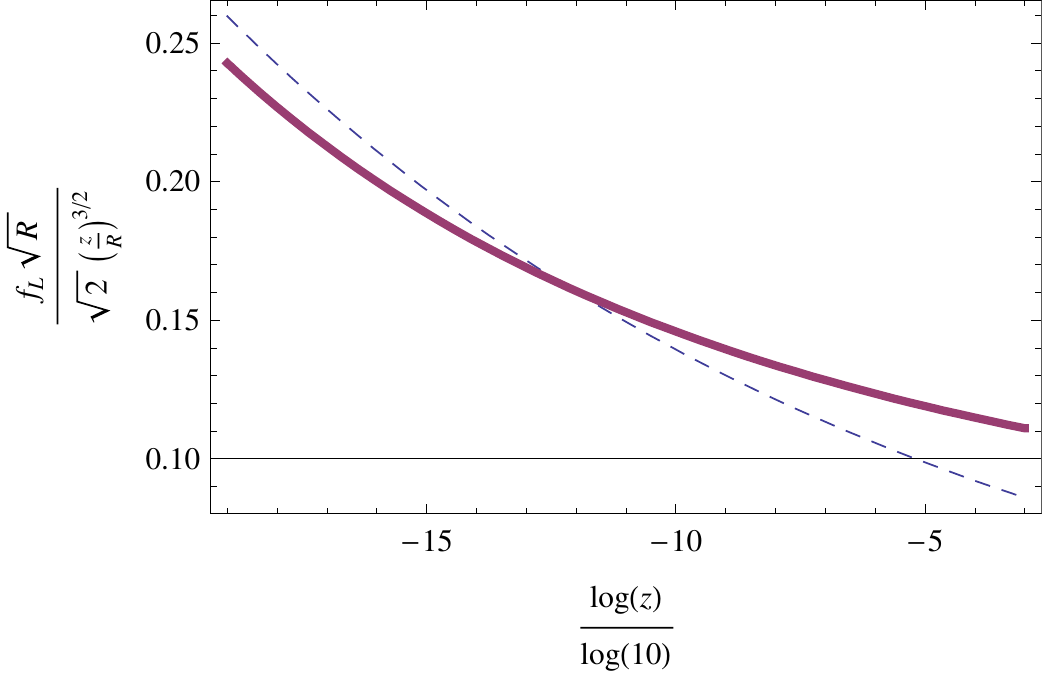}\qquad\includegraphics[width=7.5cm]{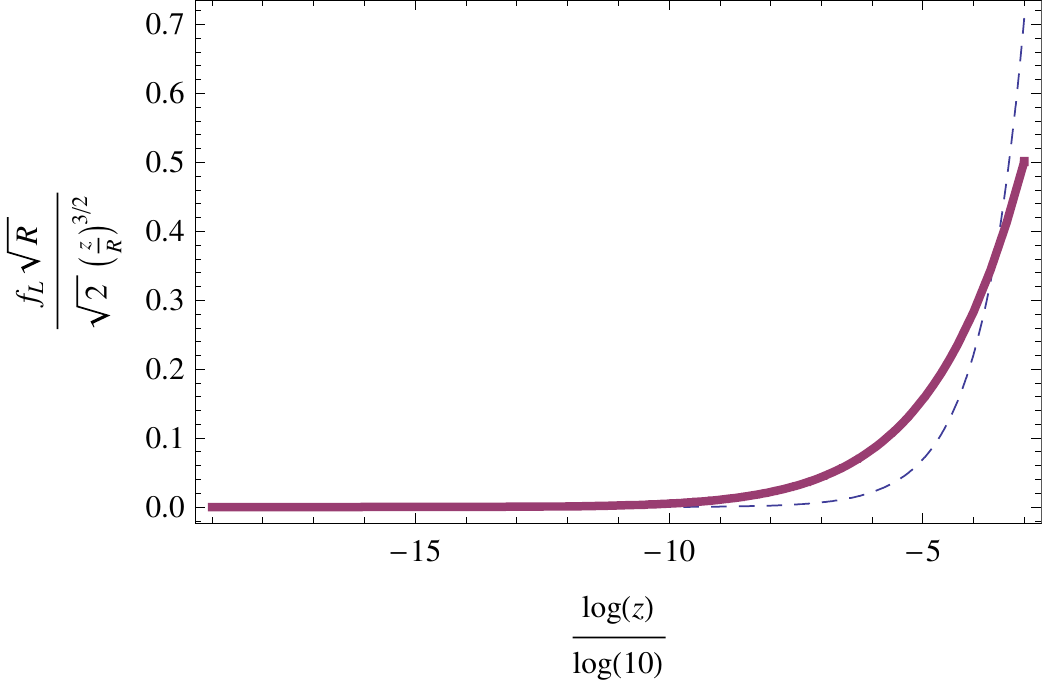}
\end{center}
\caption{Normalized zero modes. Dashed: standard bulk mass term with
$c = 1/2 + \beta$ and no scalar contribution, solid line:
 $c=1/2$ and additional scalar vev  $\phi \sim \beta \lt(\frac{R}{z}\rt)^{2 \epsilon}$.
Left Panel: $\beta=0.05$, Right panel:  $\beta=-0.506$
($\epsilon=0.01$ corresponding to $M R = 1/2$). One can read of the
value $f({c})$ of the wave function on the IR brane.
\label{zeromode}}
\end{figure}
In the limit $\epsilon \to 0$, we find the usual form of the zero
mode with a bulk mass $c+\beta$. The leading correction to the shape
of the wave-function is given by the second term and we show two
examples in Fig. \ref{zeromode}.

\end{document}